\documentclass[journal]{IEEEtran}
\IEEEoverridecommandlockouts
\usepackage{cite}
\usepackage{amsmath,amssymb,amsfonts}
\usepackage{algorithmic}
\usepackage{graphicx}
\usepackage{textcomp}
\usepackage{xcolor}
\usepackage{multirow}
\usepackage{booktabs}
\def\BibTeX{{\rm B\kern-.05em{\sc i\kern-.025em b}\kern-.08em
    T\kern-.1667em\lower.7ex\hbox{E}\kern-.125emX}}
\begin{document}

\title{FT-Pilot: Automated Fault-Tolerant RTL Rewriting via Vulnerability-Guided LLMs \\
}

\author{Weixing Liu, Zizhen Liu,~\IEEEmembership{Member,~IEEE,} 
Jing Ye,~\IEEEmembership{Member,~IEEE,}
Naixing Wang,
Cheng Liu,~\IEEEmembership{Senior Member,~IEEE,} 
Huawei Li,~\IEEEmembership{Senior Member,~IEEE,} 
and Xiaowei Li,~\IEEEmembership{Senior Member,~IEEE}
\thanks{The corresponding authors are Zizhen Liu 
and Cheng Liu.}
\thanks{W. Liu is with the School of Advanced 
Interdisciplinary Sciences, University of Chinese 
Academy of Sciences, Beijing, China. 
E-mail: liuweixing23@mails.ucas.ac.cn.}
\thanks{Z. Liu is with the State Key Lab of 
Processors, Institute of Computing Technology, 
Chinese Academy of Sciences, Beijing, China.
E-mail: liuzizhen@ict.ac.cn.}
\thanks{C. Liu and X. Li are with the State Key 
Lab of Processors, Institute of Computing 
Technology, Chinese Academy of Sciences, and 
the University of Chinese Academy of Sciences, 
Beijing, China. 
E-mail: \{liucheng, lxw\}@ict.ac.cn.}
\thanks{J. Ye and H. Li are with the State Key 
Lab of Processors, Institute of Computing 
Technology, Chinese Academy of Sciences, the 
University of Chinese Academy of Sciences, 
and CASTEST Co., Ltd. 
E-mail: \{yejing, lihuawei\}@ict.ac.cn.}
\thanks{N. Wang is with Bodesi Technologies. 
E-mail: naixingwang@bodesitech.com}

}

\maketitle

\begin{abstract}
As integrated circuit technologies continue to scale toward advanced process nodes, the continual reduction in node capacitance and supply voltage has made digital systems increasingly vulnerable to soft errors. Although traditional full-chip hardening methods can improve reliability, they often incur unacceptable area and power overhead, making selective hardening a more practical engineering solution. However, existing approaches typically rely on time-consuming fault-injection simulation to determine hardening locations through vulnerability analysis, and still depend heavily on manual strategy selection and RTL modification during the hardening stage, making them ill-suited for efficient automated reliability optimization at early design stages. To address these challenges, this paper proposes FT-Pilot, a GNN-guided LLM framework for automatic RTL soft-error hardening. The framework first employs a GNN to identify critical vulnerable assets directly at the RTL level, and then introduces an LLM-driven rewriting engine composed of an analyzer and a rewriter, which performs RTL-level fault-tolerant code rewriting with the support of dual-knowledge-base retrieval-augmented generation and an automatic repair mechanism. Experimental results show that the proposed framework can automatically generate hardened RTL designs that are syntactically correct, functionally correct, and synthesizable across multiple benchmark circuits, while significantly reducing output error rates under soft-error scenarios. This work provides a practical automated path toward shift-left reliability optimization at the RTL level.
\end{abstract}


\section{Introduction}
As semiconductor technology continues to scale toward advanced
process nodes, digital circuits are becoming increasingly sensitive
to environmental disturbances such as radiation and supply-voltage
fluctuations, making soft errors a key threat to system reliability~\cite{softerror}.
In safety-critical domains such as aerospace and autonomous driving,
even a single transient fault can lead to severe consequences,
necessitating effective soft-error protection mechanisms~\cite{iso26262}. In practice,
improving system reliability critically depends on fault-tolerant register-transfer level (RTL)
rewriting, where the original design is transformed to incorporate
appropriate protection strategies.

\begin{figure}[!t]
    \centering
    \includegraphics[width=\columnwidth]{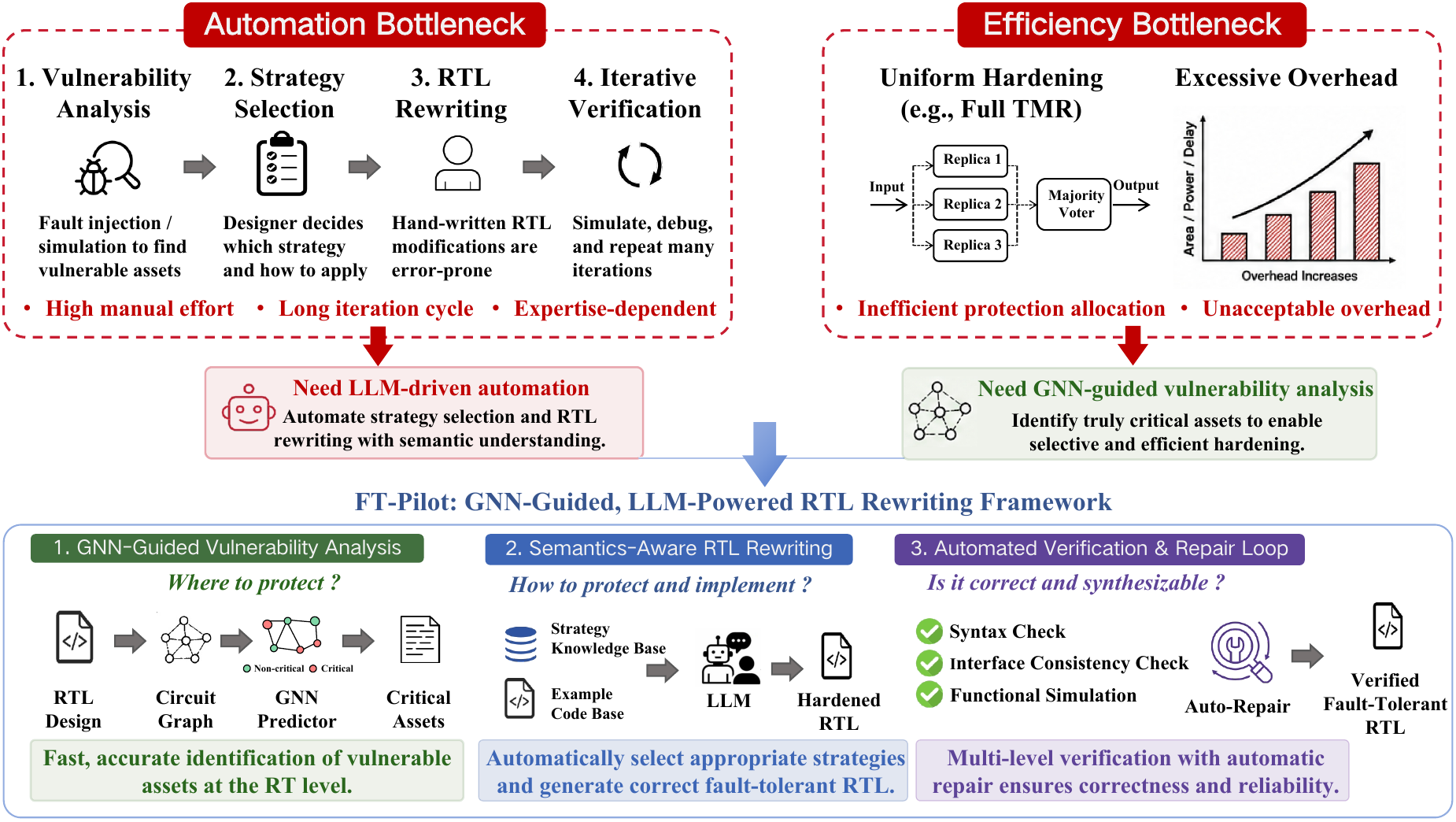}
    \caption{Motivation for FT-Pilot. Traditional RTL hardening faces two key bottlenecks: (1) an automation bottleneck due to the labor-intensive, expertise-dependent manual workflow; and (2) an efficiency bottleneck caused by uniform hardening strategies such as full TMR that incur excessive overhead. FT-Pilot addresses both challenges through GNN-guided vulnerability analysis and LLM-powered semantics-aware RTL rewriting, forming a closed-loop pipeline for selective and automated fault-tolerant hardening.}
    \label{fig:motivation}
\end{figure}

However, despite its importance, RTL-level fault-tolerant rewriting
remains largely manual. Designers are required to identify vulnerable
components, analyze fault impact, select suitable protection strategies~\cite{lyons1962tmr, hamming1950error}
(e.g., triple modular redundancy (TMR), error-correcting code (ECC), parity), and modify RTL code accordingly. This process is labor-intensive, error-prone, and difficult to scale to large
and complex designs. Existing design flows further exacerbate this
challenge. As illustrated in Fig.~\ref{fig:motivation}, vulnerability analysis is typically performed through post-synthesis gate-level fault injection~\cite{ziade2004survey}, followed by manual RTL hardening. Each iteration can take days to weeks due
to the high cost of gate-level simulation and repeated verification. More importantly, this workflow exposes a fundamental gap between
vulnerability analysis and fault-tolerant implementation. While
existing tools can identify critical components, they provide no
systematic mechanism to translate analysis results into correct and
effective RTL-level transformations. Consequently, fault-tolerant
rewriting remains ad hoc and heavily dependent on designer expertise.

This gap persists despite extensive research in electronic design automation (EDA). Existing tools provide strong support for soft-error analysis, such as fault-injection simulation~\cite{ziade2004survey} and soft-error rate (SER) estimation~\cite{zhang2006soft}, and offer limited protection mechanisms for specific structures (e.g., ECC for memories or TMR insertion~\cite{TMRG}). However, these techniques are typically restricted to predefined patterns and do not generalize to arbitrary RTL designs. Fundamentally, reliability-aware RTL rewriting is not a purely structural optimization problem, but a semantic-driven and strategy-dependent transformation task. Effective fault-tolerant modification requires understanding the functional role of design components (e.g., control logic, datapaths, and storage elements) and selecting appropriate protection strategies accordingly. Moreover, such transformations are often non-local, involving coordinated modifications across multiple modules and interfaces, which cannot be handled by localized rule-based or synthesis-driven approaches. Due to the lack of semantic understanding and the inability to perform context-aware, non-local transformations, existing EDA tools are insufficient to automate this process. As a result, RTL-level fault-tolerant rewriting remains largely manual in practice.

Recent advances in machine learning provide new opportunities to address these challenges. In particular, large language models (LLMs) have demonstrated strong capabilities in code understanding, generation, and transformation, making them promising candidates for automating RTL rewriting tasks~\cite{rtlcoder,rtlrewriter}. However, directly applying general-purpose LLMs to reliability-aware RTL hardening is fundamentally challenging. 
Although LLMs can capture code semantics, they lack domain-specific knowledge of circuit fault tolerance and protection strategy selection. Consequently, they may generate syntactically valid modifications without ensuring that the applied transformations are appropriate for soft-error mitigation. Moreover, since hardening effectiveness cannot be inferred from syntax alone, the generation process lacks reliable feedback, leading to potentially incorrect or ineffective results. Therefore, effective LLM-based RTL hardening requires explicit fault-tolerance knowledge and verification-driven guidance.

A second fundamental challenge is that fault-tolerant rewriting must be guided by structural vulnerability rather than applied uniformly. Selective hardening~\cite{selectivehardening} is necessary to balance reliability and overhead, which requires identifying vulnerable and critical assets prior to rewriting. However, such structural vulnerability information is not directly accessible to general-purpose LLMs, as it depends on circuit topology and fault-impact propagation rather than local code semantics. Existing approaches address this problem using fault-injection simulation or machine-learning models on post-synthesis gate-level netlists~\cite{Balakrishnan,lu2023,ml,lange2019ml}, which incur high analysis cost and are tightly coupled to specific technology libraries. In contrast, RTL-level representations preserve higher-level structural properties that are more stable across implementations, enabling earlier-stage and more portable vulnerability estimation. Therefore, vulnerability prediction must be integrated into the rewriting process to provide explicit structural guidance for selective RTL hardening.

To address the above challenges, this paper proposes FT-Pilot, an automated framework for reliability-aware RTL rewriting, which establishes a closed loop from vulnerability identification to fault-tolerant implementation. Specifically, the framework first converts RTL designs into And-Inverter Graph (AIG) based graph~\cite{wolf2013yosys} representations and employs a graph neural network (GNN) to identify vulnerable assets directly at the RTL level, thereby eliminating the dependence on post-synthesis gate-level netlists and enabling earlier-stage reliability analysis and optimization. Building on this foundation, we further design a two-stage LLM-driven rewriting pipeline consisting of an RTL Analyzer and an RTL Rewriter. The Analyzer is responsible for examining vulnerable assets and assigning appropriate fault-tolerance strategies, producing a structured hardening plan. Guided by this plan, the Rewriter performs strategy-aware RTL rewriting with retrieval augmentation from both a fault-tolerance semantic knowledge base and a fault-tolerance example knowledge base. Through the integration of automatic repair and multi-level verification, the proposed framework establishes an automated closed loop from vulnerability identification to RTL fault-tolerant implementation, providing a practical path toward shift-left design for reliability optimization. 
 
The main contributions of this work are as follows:
\begin{itemize}
    \item We formulate reliability-aware RTL rewriting as a unified problem that bridges vulnerability analysis and fault-tolerant implementation, and propose an automated framework that enables a closed-loop design flow for scalable soft-error hardening.

    \item  We develop an RTL-level vulnerability prediction approach that captures structurally critical assets without relying on post-synthesis gate-level representations, enabling early-stage and technology-independent reliability optimization.

    \item  We introduce a strategy-aware and knowledge-augmented RTL rewriting paradigm that integrates structured analysis and guided transformation with verification-driven refinement, improving the correctness and robustness of fault-tolerant RTL generation.
    
    \item   We implement an end-to-end soft-error hardening system and demonstrate its effectiveness on diverse benchmark circuits, achieving high prediction accuracy, stable rewriting quality, and significant reduction in soft-error vulnerability with acceptable overhead.
\end{itemize}

\section{Related Works}
\subsection{Soft Error Protection and Selective Hardening} 
A wide range of protection techniques have been proposed to mitigate soft errors such as single-event upsets (SEUs) at different abstraction levels. At the architectural level, prior work~\cite{mukherjee2005soft} primarily relies on hardware redundancy and checkpoint–rollback mechanisms, introducing redundant execution or recovery capabilities at the system level to tolerate transient faults. At the algorithm level, Huang and Abraham proposed the classical algorithm-based fault tolerance (ABFT)\cite{abft}, which provides a low-overhead approach for error detection and correction in structured computations such as matrix operations. 
At the circuit level, protection strategies are typically tailored to different types of modules. For data paths, parity checking and EDC/ECC\cite{hamming1950error} techniques are commonly employed. For control logic, robust state encoding schemes such as safe-state encoding and one-hot encoding are widely used to improve state resilience~\cite{fsm}. For memory structures, ECC remains the most mature and widely adopted protection mechanism~\cite{hamming1950error}. In addition, for timing-related soft error detection, Ernst \emph{et al.} proposed Razor\cite{razor}, which improves circuit robustness by detecting timing errors and triggering recovery mechanisms. Subsequent works, such as Bubble Razor\cite{bubblerazor}, further reduce false positives and improve practical applicability.

Among these techniques, TMR~\cite{lyons1962tmr} is widely regarded as one of the most effective hardware-level fault tolerance approaches, as it can directly mask single-point failures through triplication and majority voting. However, this robustness typically comes at the cost of over 200\% hardware overhead, along with significant penalties in area, power, and timing. To address this limitation, selective hardening techniques have been proposed~\cite{selectivehardening}. The key observation is that not all registers or circuit elements contribute equally to system-level reliability; therefore, instead of uniformly applying protection, it is more efficient to selectively harden only the most vulnerable and critical components. 
Existing approaches to selective hardening typically rely on fault injection or similar analysis techniques to identify critical nodes, followed by the application of appropriate fault tolerance strategies to balance reliability and power, performance, and area (PPA)~\cite{li2020exploring}. Traditional analysis methods such as fault injection incur substantial runtime overhead, which limits the efficiency of vulnerability assessment. Moreover, after identifying critical nodes, the subsequent steps—strategy selection and RTL-level code modification—are still largely performed manually, with limited support from automated EDA tools, resulting in a workflow with low overall automation.
\subsection{Machine Learning-Based Soft Error Vulnerability Prediction}
Traditional fault injection requires simulation to be performed exhaustively over a large number of registers, and its computational cost increases rapidly with design size. For million-gate circuits, the time overhead of such simulation-based analysis becomes prohibitively high, to the point of being impractical. To accelerate this process, Tang et al.~\cite{tang2025eraser} proposed 
a redundancy-elimination-based RTL fault simulation method. Although it improves efficiency on small- and medium-scale circuits, its benefit remains limited for large-scale designs. Researchers subsequently introduced machine learning into vulnerability analysis. Lange et al.~\cite{lange2019ml} employed machine learning models to 
predict flip-flop vulnerability and reduce fault injection cost. Lu et al. \cite{lu2023} exploited the inherently graph-structured nature of digital circuits by combining graph neural networks with dynamic behavioral features, thereby improving prediction accuracy. Das et al.~\cite{das2024gcn} further incorporated GNN interpretability 
analysis to reveal the relationship between circuit features and vulnerability, 
providing guidance for downstream hardening.

Although existing methods have achieved promising results in both analysis efficiency and prediction accuracy, they still rely on post-synthesis gate-level netlists derived from RTL. This dependence introduces two major limitations. First, vulnerability analysis can only be carried out after synthesis, making it unsuitable for rapid iteration and shift-left optimization in the early design stage. Second, the resulting prediction models are often tightly coupled to a specific technology library, so changes in the target library typically require retraining, which limits generality and portability. Therefore, there is a strong need for a method that can predict soft-error vulnerability directly at the RTL level, thereby supporting fast early-stage analysis and reliability optimization.
\subsection{Application of Large Language Models in Hardware Design Automation}
In recent years, the application of LLMs to EDA has expanded rapidly. Existing work has primarily focused on RTL generation, design space exploration, PPA optimization, verification assistance, and workflow automation. In RTL generation, systems such as AutoVCoder \cite{autovcoder}, RTLCoder\cite{rtlcoder}, and MAGE~\cite{mage} have shown that domain-specific fine-tuning, retrieval-augmented generation (RAG), and role-based collaboration can produce strong results on Verilog generation tasks. In design space exploration, LLM4GV~\cite{llm4gv} explores the design 
space of GEMM under performance constraints, while Yao et al.~\cite{rtlrewriter} 
demonstrate that LLMs can approach human-expert performance in RTL-level PPA optimization through retrieval-augmented rewriting. In verification and debugging, AssertionLLM~\cite{assertion} and Self-HWDebug~\cite{selfhwdebug} demonstrate the usefulness of LLMs for assertion generation and hardware debugging. Beyond RTL-level tasks, HLSPilot~\cite{hlspilot} further extends LLM-based automation to the HLS abstraction layer, leveraging RAG-based strategy retrieval to transform sequential C/C++ into optimized FPGA accelerators. In addition, some recent studies~\cite{chateda} have extended LLMs into agent-style systems for EDA workflows, supporting task decomposition, script generation, and tool invocation.

Overall, existing LLM-for-EDA research has concentrated on code generation, optimization, and verification, while RTL rewriting for soft-error-tolerant hardening remains largely unexplored. Compared with general RTL generation tasks, fault-tolerant rewriting requires not only syntactic correctness and functional preservation, but also identification of protection targets, selection of appropriate hardening strategies, and careful trade-offs between reliability improvement and implementation overhead. These requirements place substantially stronger demands on consistency and verifiability, and remain insufficiently studied.
\section{Methodology}
\begin{figure*}[t!]
    \centering
    \includegraphics[width=\textwidth]{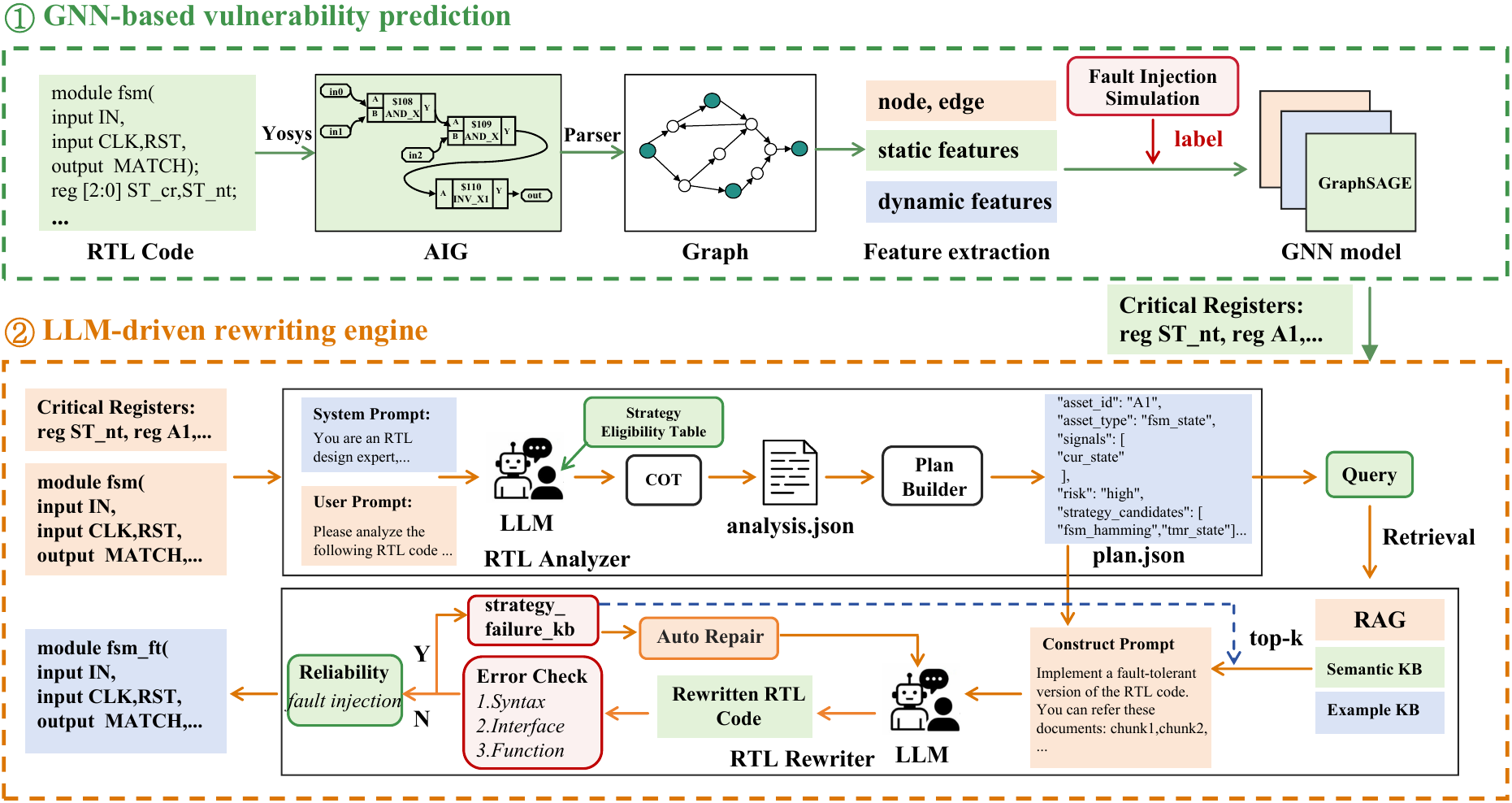}
    \caption{Overview of the proposed FT-Pilot framework for RTL soft error hardening.}
    \label{fig:framework}
\end{figure*}

\subsection{Proposed Framework Overview}\label{AA}
The goal of this work is to automate fault-tolerant RTL rewriting,
where a given RTL design is transformed into a resilient version
through structure-aware and strategy-driven code modifications.
As illustrated in Fig.~\ref{fig:framework}, FT-Pilot establishes a
closed-loop rewriting pipeline that integrates structural
vulnerability guidance, LLM-based transformation, and
verification-driven refinement.

A key challenge in fault-tolerant RTL rewriting is that effective
modification must be guided by circuit-level vulnerability rather
than applied uniformly. To address this, we first derive structural
vulnerability information directly at the RTL level. Specifically,
the input RTL design is converted into an AIG-based graph
representation, from which both structural and activity-related
features are extracted. A GNN is then used
to predict vulnerable assets (e.g., critical registers), providing
explicit structural guidance for subsequent rewriting. Unlike
conventional approaches that rely on expensive post-synthesis
analysis, this step enables efficient and early-stage identification
of protection targets.

Building on the vulnerability guidance, we construct a structured
analysis-and-transformation pipeline for strategy-aware RTL rewriting.
The pipeline consists of an RTL Analyzer and an RTL Rewriter, where
high-level rewriting decisions are first derived and then
systematically applied to generate fault-tolerant RTL.
Given the original RTL code and the identified vulnerable assets,
an RTL Analyzer first interprets their functional roles and assigns
candidate fault-tolerance strategies based on a predefined strategy
eligibility table. The analysis results are organized into a
structured representation, which is further processed to generate
a high-level rewriting plan describing target assets, risk levels,
and candidate protection strategies.

To ensure both correctness and effectiveness, we integrate
knowledge-augmented generation with verification-driven refinement
into a unified rewriting process. Specifically, we adopt a RAG~\cite{rag} mechanism to incorporate both
semantic knowledge of fault-tolerance techniques and representative
transformation examples, improving the quality and reliability of
LLM-based code generation. Guided by the rewriting plan and retrieved
knowledge, the RTL Rewriter performs strategy-aware transformation
to generate fault-tolerant RTL. The generated design is then subjected to multi-level verification,
including syntax, interface, and functional correctness checks, along
with fault-injection-based reliability evaluation. If violations or
ineffective transformations are detected, an automatic repair module
is triggered to iteratively refine the code. Through this tightly
coupled loop of knowledge-guided generation and verification-driven
feedback, the proposed framework ensures high-quality fault-tolerant
RTL rewriting with minimal manual effort.

The rest of this section is organized as follows. Section~III-B presents the RTL-level vulnerability prediction method, and Section~III-C describes the strategy-aware LLM-based RTL rewriting process.

\subsection{GNN-Based Vulnerability Prediction}

\begin{figure}[htbp]
    \centering
    \includegraphics[width=\columnwidth]{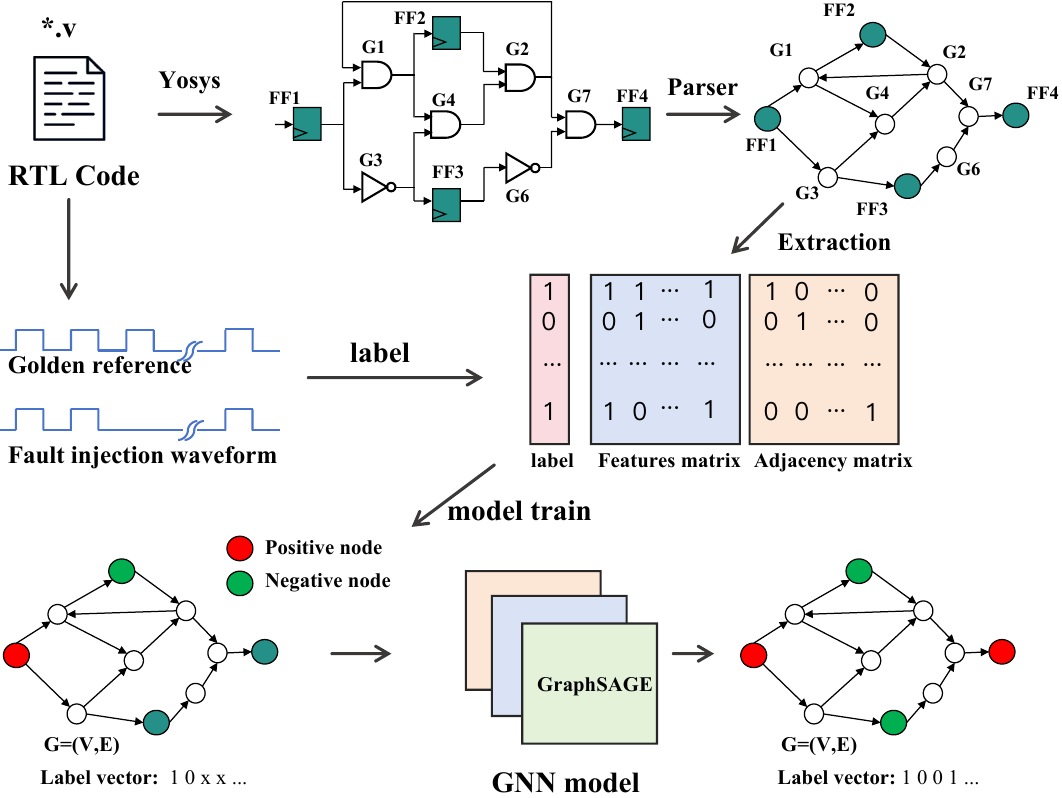}
    \caption{Architecture of the GNN-based vulnerability prediction module (Stage 1).}
    \label{fig:gnn_framework}
\end{figure}

To perform RTL-level vulnerability prediction, we first construct a circuit graph representation and extract the corresponding features, then generate supervision labels through fault injection, and finally employ a GNN model to perform register-level vulnerability prediction. The overall pipeline is illustrated in Fig.~\ref{fig:gnn_framework}.

\textbf{1) Graph Construction and Representation.}
To enable GNN-based vulnerability prediction at the RTL level, we first convert the RTL design into a graph representation. Specifically, we adopt the AIG as an intermediate representation and construct the input graph for the GNN on top of it. Compared with post-synthesis gate-level netlists, AIG provides a more uniform representation and is decoupled from specific technology libraries. It also avoids the additional cost of synthesis and technology mapping, making it more suitable for reliability analysis in the early design stage.

\begin{figure}[htbp]
    \centering
    \includegraphics[width=\columnwidth]{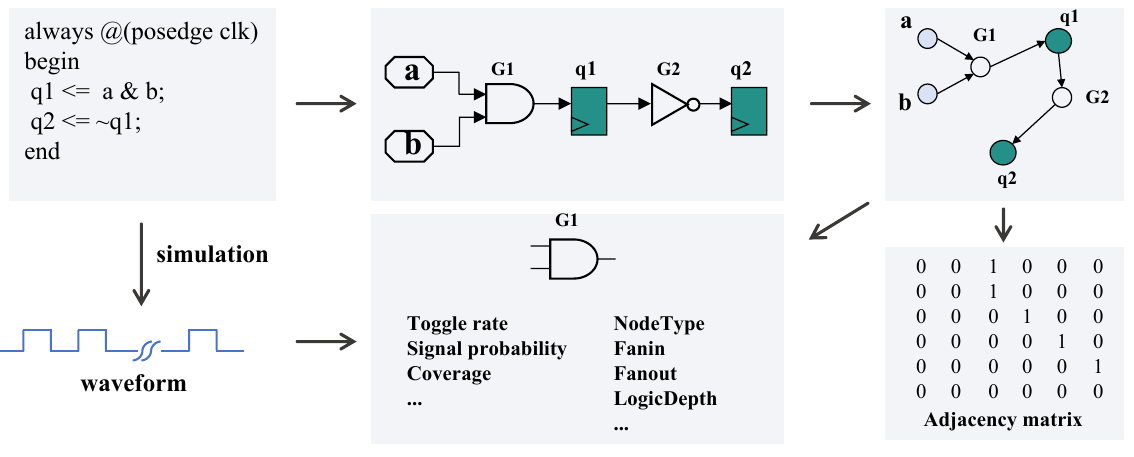}
    \caption{Illustration of graph construction and feature extraction from an RTL design.}
    \label{fig:rtl-graph}
\end{figure}
For a given RTL design, we first use the open-source synthesis tool Yosys to translate it into an AIG representation~\cite{wolf2013yosys}. In this representation, logic gates and flip-flops are modeled as graph nodes, while signal connections between them are modeled as graph edges. Fig.~\ref{fig:rtl-graph} shows a simple example. Based on the parsed AIG, we further construct a directed graph
\[
G = (V, E),
\]
where the node set $V$ includes primary inputs (PIs), primary outputs (POs), logic gates, and flip-flops, and the edge set $E$ represents the logical connectivity between nodes. Accordingly, the graph is represented by a node feature matrix
\[
X \in \mathbb{R}^{N \times F}
\]
and an adjacency matrix
\[
A \in \mathbb{R}^{N \times N},
\]
where $N$ denotes the total number of nodes and $F$ denotes the feature dimension.

In addition, to support efficient vulnerability back-annotation at the RTL level, we explicitly establish the correspondence between register nodes in the AIG and register objects in the RTL source code during graph construction. As a result, once vulnerability prediction has been performed on the register nodes in the graph, the predicted results can be directly mapped back to the RTL level, thereby producing a vulnerability report for critical registers in the RTL design. Since this correspondence is established during preprocessing, the back-annotation process introduces almost no additional overhead.

\textbf{2) Feature Extraction.}
After graph construction and representation, we further extract feature representations for graph nodes for vulnerability prediction. Since node vulnerability depends not only on the topological position of a node in the circuit, but also on its dynamic behavior during operation, we jointly consider static topological features and dynamic temporal features~\cite{features} under representative workloads for register vulnerability modeling. For a graph node $v_i$, its final feature vector is defined as
\[
\mathbf{x}_i = [\mathbf{x}_i^{(s)} \, \| \, \mathbf{x}_i^{(d)}],
\]
where $\mathbf{x}_i^{(s)}$ denotes the static topological features, $\mathbf{x}_i^{(d)}$ denotes the dynamic behavioral features, and $\|$ represents vector concatenation.

The static topological features are used to characterize the structural position of a node and its local connectivity in the circuit. Specifically, we extract the following features:
\begin{itemize}
    \item \textbf{Node type}: a one-hot encoding indicating different node categories, including primary inputs, primary outputs, logic gates, and flip-flops.
    \item \textbf{Fan-in/fan-out counts}: the numbers of incoming and outgoing edges of a node, which reflect its local connectivity complexity and its influence on upstream and downstream logic.
    \item \textbf{Logic depth}: including forward logic depth and backward logic depth, which represent the logical distances from the current node to the primary input nodes and primary output nodes, respectively, thereby characterizing its relative position along the signal propagation path.
    \item \textbf{Neighbor-type distribution}: the type distribution of one-hop neighbors, which describes the structural context around the node.
\end{itemize}

The dynamic behavioral features are used to capture node activity under representative workloads. Specifically, we collect temporal behavior information through RTL simulation and extract the following dynamic features:
\begin{itemize}
    \item \textbf{Toggle rate}: the number of output signal transitions of a node during simulation; a higher toggle rate indicates more frequent state changes.
    \item \textbf{Signal probability}: the fraction of time for which the node output remains at logic high over the total execution time.
    \item \textbf{Coverage}: a measure of how fully a node signal is activated during simulation. If a node signal takes both logic 0 and logic 1 during simulation, its coverage is set to 1; if the signal remains constant (either always 0 or always 1), its coverage is set to 0.5.
\end{itemize}

By jointly modeling static topological features and dynamic behavioral features, we provide the subsequent GNN with a more comprehensive node representation. This allows the model to capture not only the structural position of registers in the circuit, but also their runtime activity patterns, thereby enabling more accurate vulnerability prediction.

\textbf{3) Label Generation.}
We generate vulnerability labels for the training dataset using conventional fault-injection simulation. Specifically, we use Synopsys Z01X~\cite{synopsys_z01x} as the fault-injection tool to inject SEU faults into circuit registers and observe whether the faults lead to erroneous outputs at the primary outputs of the design, thereby assessing the impact of each register on overall system reliability. For each RTL design, we prepare a representative workload testbench. We first perform fault-free simulation and record the resulting output waveform as the golden reference. Then, over the full execution window of the circuit, we inject a single bit-flip fault into each register at a random time point. After fault injection, the faulty output waveform is collected and compared against the golden reference. If the two waveforms differ, the injected fault is considered to have caused an output error.

Based on this procedure, we perform fault-injection simulation on every register in each RTL design and compute its Architectural Vulnerability Factor (AVF). For a register $r_i$, the AVF is defined as the ratio between the number of fault injections that lead to output errors and the total number of injections, i.e.,
\[
\mathrm{AVF}(r_i)=\frac{N_i^{\mathrm{err}}}{N_i^{\mathrm{inj}}},
\]
where $N_i^{\mathrm{err}}$ denotes the number of injections on register $r_i$ that result in output errors, and $N_i^{\mathrm{inj}}$ denotes the total number of injections on that register. A higher AVF indicates that faults occurring in the register are more likely to propagate to the primary outputs and thus pose a greater threat to system reliability. To account for different reliability requirements across tasks, 
we predefine a per-design AVF threshold based on the distribution 
of register AVF values in each circuit, and label registers above 
this threshold as vulnerable (label~1) and the rest as 
non-vulnerable (label~0). In this way, register vulnerability 
prediction is formulated as a binary classification problem.

\textbf{4) Graph Neural Networks.}
We adopt GraphSAGE~\cite{graphsage} as the vulnerability prediction model. Compared with conventional GCN, which performs convolution over the entire graph, GraphSAGE learns node representations through neighborhood sampling and feature aggregation, making it more suitable for large-scale circuit graphs. The core idea is as follows: for a target node $v$, its neighborhood set $\mathcal{N}(v)$ is sampled, and the neighbor features are aggregated to update the embedding of the target node. The node representation at the $k$-th layer is updated as
\[
h_v^{(k)} = \sigma\left(W^{(k)} \cdot \mathrm{AGG}\left(\left\{h_u^{(k-1)}, \forall u \in \mathcal{N}(v)\right\}\right)\right)
\]
where $\mathrm{AGG}(\cdot)$ denotes the aggregation function (mean aggregation is used in this work), $W^{(k)}$ is the learnable weight matrix, and $\sigma$ is the ReLU activation function.

\begin{figure}[htbp]
\centerline{\includegraphics[width=\columnwidth]{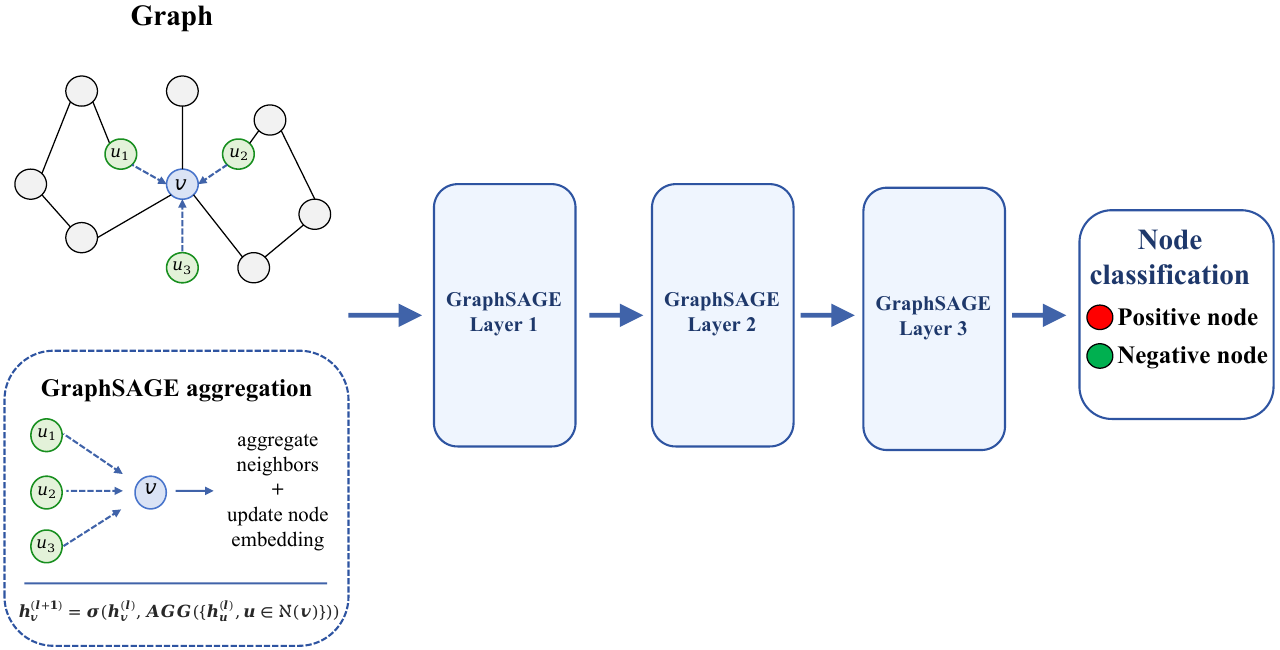}}
\caption{Architecture of the three-layer GraphSAGE network.}
\label{graphsage}
\end{figure}

As shown in Fig.~\ref{graphsage}, the proposed network consists of three SAGEConv layers. The first two layers map node features from the input dimension to the hidden space, each followed by BatchNorm, ReLU activation, and Dropout to accelerate convergence and mitigate overfitting. The third layer maps the hidden features to a two-dimensional output space and directly outputs the logits.

The model is trained using a class-weighted cross-entropy loss to address the class imbalance between vulnerable and non-vulnerable registers:
\[
\mathcal{L} = -\frac{1}{N}\sum_{i=1}^{N} w_{y_i} \left[y_i \log(\hat{p}_1^{(i)}) + (1-y_i)\log(1-\hat{p}_1^{(i)})\right]
\]
where $w_{y_i}$ denotes the class weight, and $\hat{p}_1^{(i)}$ denotes the Softmax-normalized probability that node $i$ is predicted to be vulnerable. During inference, $\arg\max$ is used to obtain the final predicted class.

\subsection{LLM-Driven Rewriting Engine}
Building on the vulnerability prediction results from the previous stage, we design an LLM-driven rewriting engine to automatically harden the RTL code against soft errors. As illustrated in the lower half of Fig.~\ref{fig:framework}, the engine comprises two submodules: the RTL Analyzer and the RTL Rewriter. The RTL Analyzer examines each critical asset by jointly considering the vulnerability report and the RTL code semantics, assigns candidate fault-tolerance strategies, and produces a structured hardening plan. The RTL Rewriter then carries out the actual code transformation according to this plan through a retrieval-augmented generation mechanism. The following subsections detail the design of each submodule.
\begin{figure}[htbp]
\centerline{\includegraphics[width=\columnwidth]{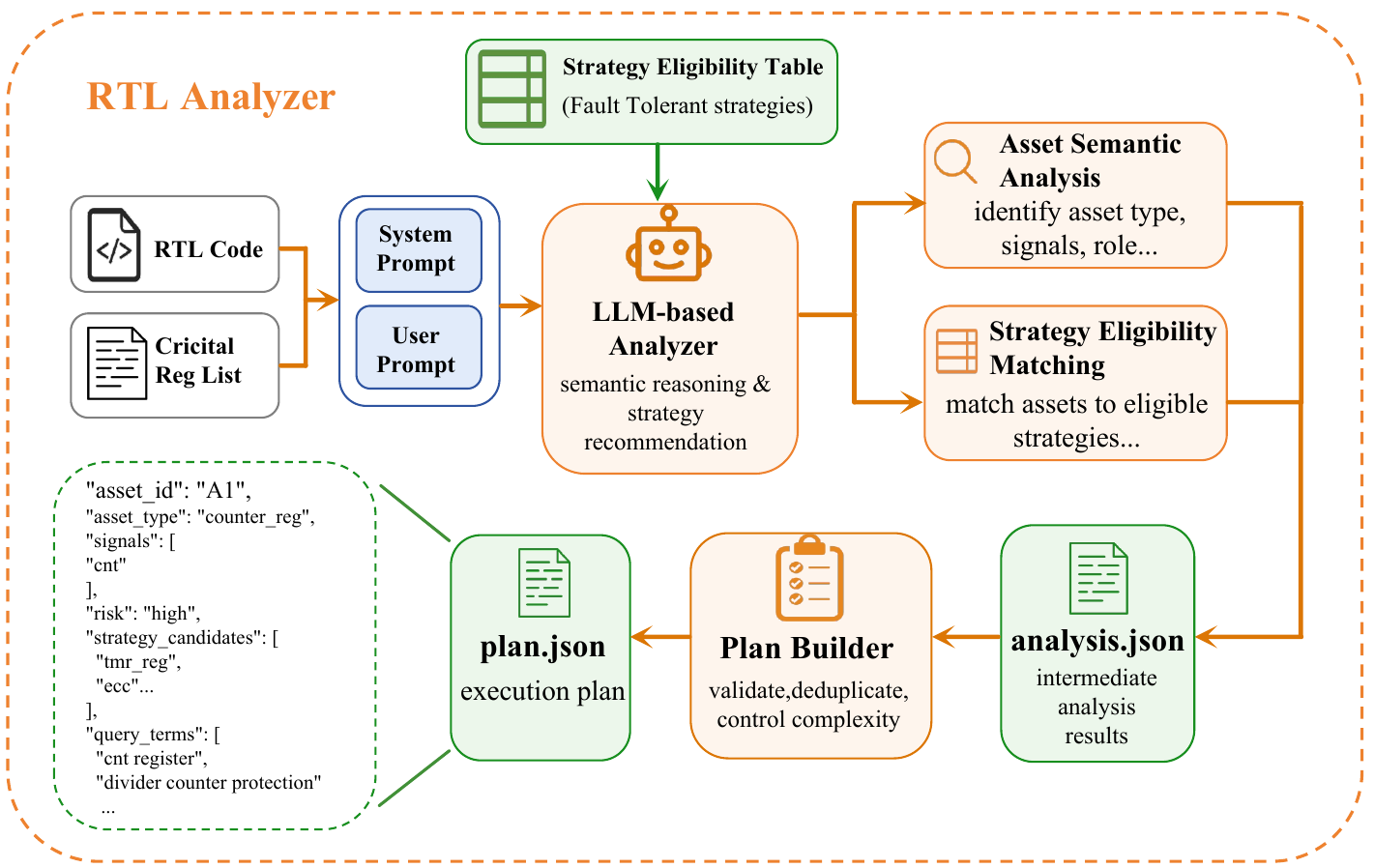}}
\caption{Detailed workflow of the RTL Analyzer module.}
\label{rtlanalyzer}
\end{figure}

\textbf{1) RTL Analyzer.}
The RTL Analyzer bridges vulnerability identification and RTL rewriting, transforming the critical vulnerable assets identified upstream into an actionable hardening plan. As illustrated in Fig.~\ref{rtlanalyzer}, it takes the vulnerability report and the original RTL code as input, leverages code-level semantic analysis to understand the functional role of each asset within the circuit, and formulates appropriate fault-tolerance strategies accordingly. The overall process proceeds through three stages: LLM-based semantic analysis, strategy eligibility matching, and plan validation, ultimately producing a structured execution plan in JSON format.

To improve the stability and controllability of the LLM when analyzing RTL code, we adopt a chain-of-thought (CoT)~\cite{cot} prompting approach that decomposes the analysis into a sequence of reasoning steps. The model first identifies the overall structural category of the module—for instance, a finite state machine (FSM), a counter, a datapath, or a storage block. It then examines each critical asset by considering the signal's definition, update pattern, control dependencies, and usage sites to determine the asset's type, bit-width, and functional semantics. Finally, guided by a predefined Strategy Eligibility Table, the model selects semantically matched candidate fault-tolerance strategies for each asset.

The Strategy Eligibility Table serves as an explicit constraint mechanism introduced to curb the inherent randomness of the LLM during strategy selection and to prevent the assignment of strategies incompatible with a given asset type. As shown in Table~\ref{tab:eligibility}, the table enumerates the set of legitimate fault-tolerance strategies for each asset category, restricting the model to choose only within a semantically appropriate range. For example, Hamming encoding and TMR are among the eligible strategies for FSM state registers, whereas ECC and parity checking are better suited for wide data registers. Beyond eligibility, we further incorporate implementation cost considerations under different scenarios. For wide registers, for instance, ECC typically incurs lower overhead than TMR, while the opposite may hold for narrow registers. Supplying such prior knowledge enables the Analyzer not only to judge whether a strategy is applicable but also to make a more informed selection among multiple legitimate candidates.

Because LLM outputs are inherently open-ended and stochastic, the candidate assets and their strategy assignments produced by the preceding analysis do not always satisfy the predefined constraints. The model may, for instance, deviate from the Strategy Eligibility Table or assign multiple strategies to the same asset. Passing such results directly to the downstream Rewriter could lead to over-hardening, redundant protection, and prompt bloating. To address this, we introduce a Plan Builder module within the RTL Analyzer that applies deterministic rules to validate and refine the analysis results before they are emitted. Specifically, it performs three checks. First, a static structure check parses the sensitivity lists of \texttt{always} blocks to identify signals that are assigned only in combinational logic; such signals correspond to no physical flip-flop, and applying register-level hardening to them would cause synthesis errors, so they are removed from the candidate set. Second, a deduplication step merges overlapping assets that refer to the same physical register, eliminating redundant protection. Third, a strategy compliance check verifies each assigned strategy against the Strategy Eligibility Table and automatically replaces any non-compliant assignment with the preferred strategy for that asset type.

After the above analysis and validation, the RTL Analyzer outputs the hardening plan in a structured JSON format. Each entry records the name, type, and recommended fault-tolerance strategy of the asset to be protected, along with query keywords for subsequent RAG retrieval, thereby providing the downstream RTL Rewriter with well-defined rewriting directives.

\begin{table}[htbp]
\caption{Strategy Eligibility Table}
\begin{center}
\begin{tabular}{|c|l|l|}
\hline
\textbf{\#} & \textbf{Asset Type} & \textbf{Applicable Strategies} \\
\hline
1 & fsm\_state & \checkmark~tmr\_state, \checkmark~fsm\_hamming, \\
  &            & $\circ$~one\_hot, $\circ$~illegal\_detect, $\circ$~parity \\
\hline
2 & counter\_reg & \checkmark~tmr\_reg, \checkmark~hamming, \\
  &              & $\circ$~cnt\_comp, $\circ$~secded, $\circ$~parity \\
\hline
3 & datapath\_reg & \checkmark~tmr\_reg, \checkmark~hamming, \\
  &               & $\circ$~secded, $\circ$~parity, $\circ$~parity\_byte \\
\hline
4 & control\_reg & \checkmark~tmr\_reg, $\circ$~parity, $\circ$~watchdog \\
\hline
5 & memory & \checkmark~ecc\_fifo, \checkmark~sram\_ecc, \\
  &        & $\circ$~scrubbing, $\circ$~gray\_ptr, $\circ$~parity\_byte \\
\hline
6 & matrix\_unit & \checkmark~abft\_sec, \checkmark~abft, $\circ$~checksum \\
\hline
\multicolumn{3}{l}{\footnotesize \textbf{Legend}: \checkmark~Preferred (correction-capable) $\cdot$ $\circ$~Alternative} \\
\multicolumn{3}{l}{\footnotesize (detection-only or high overhead)}
\end{tabular}
\label{tab:eligibility}
\end{center}
\end{table}

\begin{figure}[t]
\centering
\includegraphics[width=\columnwidth]{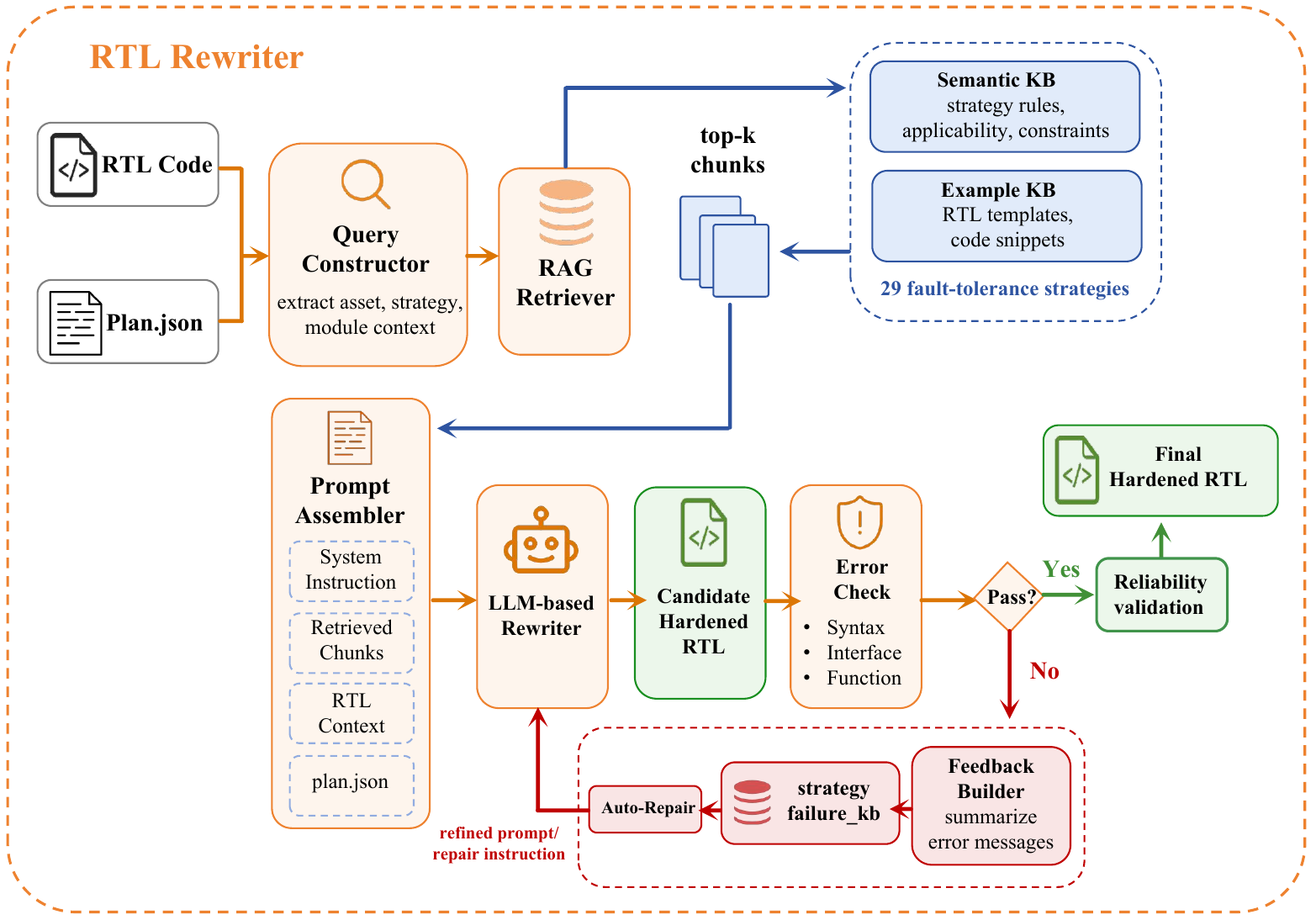}
\caption{Detailed workflow of the RTL Rewriter module.}
\label{fig:rewriter}
\end{figure}

\textbf{2) RTL Rewriter.}
RTL Rewriter is the core component of the rewriting engine and is responsible for generating the hardened RTL code according to the rewriting plan produced in the previous stage. This component adopts a RAG scheme to retrieve information relevant to the current rewriting task from a prebuilt fault-tolerance strategy knowledge base, thereby assisting RTL rewriting. In addition, we introduce an Auto-Repair mechanism, which can automatically trigger re-rewriting when the LLM produces erroneous code based on the detected error messages. The combination of these mechanisms effectively alleviates the hallucination problem of large language models in this domain-specific task. Fig.~\ref{fig:rewriter} illustrates the overall architecture of RTL Rewriter. The implementation details are described as follows.

\textbf{Dual knowledge-base architecture.} To enable the LLM to accurately and reliably exploit domain knowledge in fault-tolerant digital integrated circuit design, and to further improve its practicality and trustworthiness, we construct two one-to-one aligned knowledge bases: a fault-tolerance strategy semantic knowledge base (Semantic KB) and an example knowledge base (Examples KB). The Semantic KB contains formal descriptions of 29 fault-tolerance strategies, covering categories such as register protection, FSM hardening, datapath encoding, memory ECC, and system-level monitoring. Each knowledge entry records structured information including the applicable scenarios, implementation principles, interface constraints, and overhead estimation of the corresponding strategy. The Examples KB stores manually checked RTL rewriting templates corresponding to these semantic strategies. The cooperation between the Semantic KB and the Examples KB significantly improves the correctness of the generated code.

\textbf{Retrieval-augmented generation.} We use ChromaDB as the vector database to store both the semantic knowledge and the code examples of fault-tolerance strategies, and use a text embedding model to map the documents into a high-dimensional vector space. During code generation, the system constructs a retrieval query according to the type of the current asset, the selected strategy, and the query keywords provided by the Plan Builder. The same embedding model is then used to encode the query into a vector, after which the system retrieves the top-$k$ most relevant strategy descriptions using cosine similarity. Based on the retrieved strategy identifiers, the system further extracts the corresponding RTL code templates directly from the Examples KB. Finally, the retrieved results, together with the original RTL code, asset information, and rewriting plan, are assembled into a structured prompt and fed to the LLM to perform the RTL rewriting task.

\textbf{Context assembly.} The retrieved knowledge is assembled together with other contextual information into a structured prompt before being sent to the LLM. Specifically, the prompt includes: (i) the original RTL code and the hardening plan; (ii) the retrieved strategy descriptions and RTL code templates; (iii) an asset summary, including the signal list, asset type, and assigned strategy; and (iv) the rationale behind strategy selection. This structured prompt provides both ``what to do'' and ``how to do it'': the former is specified by the rewriting plan and the strategy rationale, while the latter is provided by the retrieved code templates. In this way, the generation space of the LLM is significantly constrained, reducing the probability of hardware-specific hallucinations.

\textbf{Collection and use of failure knowledge.} To further improve rewriting robustness, we maintain a failure knowledge base that is accumulated across runs. Whenever an error is detected during verification, the system categorizes the error message into a predefined error type and records it in the knowledge base using the pair (strategy name, error type) as the key. When the same strategy is selected again in a later run, the system retrieves the corresponding historical failure patterns and injects them into the prompt as warnings, enabling the LLM to proactively avoid common error types.

\textbf{Auto-Repair mechanism.} Even with RAG assistance, the code generated by the LLM may still contain errors such as port mismatches and multi-driver conflicts. To address this, we design an Auto-Repair module that employs an iterative repair loop. The generated code first undergoes multi-level verification, which sequentially checks syntactic correctness, synthesizability, interface consistency, and functional correctness using a combination of compilation, synthesis, and functional simulation tools to validate the rewritten design from different perspectives. If any check detects an error, an error classifier maps the raw error message to a predefined error category, each associated with a structured repair directive. The system then assembles a repair prompt comprising the original RTL, the faulty code version, the classified error information, and a condensed RAG context, guiding the LLM to correct the issue in the next generation round. Failed cases are recorded in the Strategy Failure KB to accumulate error knowledge for future reference. This loop executes for at most three rounds to strike a balance between repair success rate and computational cost. Code that passes all verification stages is further subjected to fault injection testing to confirm that it achieves the expected reliability improvement under soft error scenarios.

\section{Experiments}
\subsection{Experiment Setup}
To evaluate the effectiveness of the proposed framework, we conduct experiments on 14 open-source RTL designs collected from the RTLLM benchmark suite~\cite{rtllm_benchmark}, the OpenCores IP repository~\cite{opencores}, and public GitHub repositories. These designs cover a variety of representative application scenarios, including arithmetic computation, control logic, datapath processing, bus interfaces, and memory structures, and their full list is given in Table~\ref{tab:benchmark}. In terms of design size, the benchmarks span three scales: small, medium, and large. Small designs are approximately 100 lines of code, medium designs are around 500 lines, and large designs are multi-module systems ranging from about 1{,}000 to 3{,}000 lines of code. The large-scale benchmarks include the mriscv processor core~\cite{mriscv} as well as the i2c\_master and spi\_master bus controllers. Among them, mriscv is a complete RV32I processor core composed of nine submodules.

We employ two complementary benchmark sets that target different evaluation goals. The GNN vulnerability predictor is evaluated on four medium-to-large circuits (mriscv, SPI, I2C, XGE-MAC) that provide sufficient register populations for meaningful classification metrics. The LLM rewriting engine is evaluated on a broader set of 14 designs spanning small to large scales (Table~\ref{tab:benchmark}), all of which fit within current LLM context limits. Scaling LLM-based rewriting to designs with thousands of registers, such as XGE-MAC, remains an open challenge that we discuss further in Section~V-C.

\begin{table}[htbp]
\centering
\caption{Benchmark suite used in our experiments.}
\label{tab:benchmark}
\small
\begin{tabular}{llrcc}
\toprule
\textbf{Scale} & \textbf{Design} & \textbf{LoC} & \textbf{Category} & \textbf{\#Module} \\
\midrule
\multirow{6}{*}{Small}
  & fsm1            & 128  & Control    & 1 \\
  & alu             &  87  & Arithmetic & 1 \\
  & fifo            & 202  & Storage    & 1 \\
  & parallel2serial &  63  & Datapath   & 1 \\
  & serial2parallel & 107  & Datapath   & 1 \\
  & memory          & 146  & Storage    & 1 \\
\midrule
\multirow{5}{*}{Medium}
  & crc             & 261  & Algorithm  & 1 \\
  & serial\_io      & 284  & Interface  & 1 \\
  & bus\_a          & 572  & Interface  & 1 \\
  & gemm            & 415  & Arithmetic & 1 \\
  & uart            & 360  & Interface  & 1 \\
\midrule
\multirow{3}{*}{Large}
  & i2c\_master     & 1650 & Interface  & 4 \\
  & spi\_master     & 1089 & Interface  & 3 \\
  & mriscv          & 2665 & Processor  & 9 \\
\bottomrule
\end{tabular}
\end{table}

The performance of the GNN-based vulnerability prediction stage is evaluated on four designs, all of which are obtained from OpenCores. To construct the dataset, we first use Yosys to convert these circuits into AIG graph representations and extract their static features. We then develop typical workload testbenches for simulation to obtain the corresponding dynamic features. In addition, we employ the Synopsys Z01X fault injection tool to conduct fault injection experiments under the same representative workloads. Based on the resulting fault injection data, we construct a complete dataset for each design. For every design, the dataset is split into 70\% for training and 30\% for validation. We use Accuracy, Precision, and Recall as the primary evaluation metrics to measure the classification performance of the model in identifying critical vulnerable registers, and compare the prediction results against the recent method proposed by Lu et al.~\cite{lu2023}. Our GNN model is implemented with PyTorch Geometric~\cite{pytorchgeometric} and adopts a three-layer GraphSAGE architecture. The input feature dimension is 19, the hidden dimension is 64, and the output dimension is 2, corresponding to the two classes, vulnerable and non-vulnerable.

The effectiveness of the LLM-based rewriting stage is evaluated on four aspects: syntactic correctness, synthesizability, reliability under fault injection, and hardware overhead. We compare the proposed framework against three baselines, namely Unhardened, Full TMR, and GNN+TMRG. Here, Unhardened denotes the original RTL design without protection and serves as a reference for both area and reliability; Full TMR applies triple modular redundancy uniformly to all registers, representing a global hardening scheme; GNN+TMRG first identifies vulnerable registers using a GNN predictor and then applies TMR using the TMRG tool~\cite{TMRG}. TMRG is an open-source RTL-level tool that automates TMR insertion by triplicating target 
registers and adding majority voters. FT-Pilot denotes the proposed GNN-guided LLM-based rewriting framework, which selectively applies a range of fault-tolerance strategies, including TMR and Hamming-based SEC codes. All four configurations are evaluated on the same benchmark to ensure a fair comparison.

The rewriting module adopts Claude Opus 4.6 as the primary model and is further compared with GPT-5.3, GLM-5, and Qwen-3.6 Plus to assess sensitivity to the underlying LLM. The retrieval-augmented generation component is implemented using ChromaDB, where the knowledge base is embedded using the Qwen-text-embedding-v4 model, and the top-3 relevant strategy entries are retrieved for each query. The Auto-Repair mechanism performs up to three iterations of refinement.

To evaluate the quality of LLM-generated RTL code, we adopt the widely used pass@k metric~\cite{passk} from code generation. For each design, $n=10$ candidate outputs are generated independently using sampling parameters $T=0.8$ and $\textit{top\_p}=0.95$. Let $c$ denote the number of candidates that pass functional verification. The pass@k metric is computed as:
\[
\mathrm{pass@}k = \mathbb{E}\left[1 - \frac{\binom{n-c}{k}}{\binom{n}{k}}\right]
\]
which represents the probability that at least one correct sample is obtained when drawing $k$ samples without replacement from the $n$ candidates. We report both \textit{pass@1} and \textit{pass@3}, where \textit{pass@1} reflects single-shot success, and \textit{pass@3} captures the success rate under a small number of retries.

For the rewritten RTL code, we use iverilog~\cite{williams2002iverilog} for syntax checking and ModelSim together with the original benchmark testbenches for functional verification. The original testbenches are reused to preserve behavioral equivalence under representative workloads. Reliability improvement is evaluated through fault injection using Synopsys Z01X. In addition, hardware overhead is analyzed by synthesizing the rewritten designs with Yosys and the Nangate45 open-source technology library~\cite{stine2007nangate}.
\subsection{GNN Vulnerability Prediction Results}
In this section, we first evaluate the classification performance of the GNN-based vulnerability predictor, since its prediction accuracy directly affects the rationality of hardening target selection in the downstream LLM rewriting stage. We conduct experiments on four representative designs (mriscv, SPI, I2C, and XGE-MAC) to assess the proposed RTL-level GraphSAGE-based critical flip-flop prediction model, and compare the results with those of the state-of-the-art netlist-level prediction method proposed by Lu et al.\cite{lu2023}. The results are summarized in Table~\ref{tab:gnn-main}.
\begin{table}[htbp]
\centering
\caption{Vulnerability prediction performance (\%). 
``---'' indicates metrics not reported in~\cite{lu2023}.}
\label{tab:gnn-main}
\small
\setlength{\tabcolsep}{4pt}
\begin{tabular}{llcccc}
\toprule
\textbf{Design} & \textbf{Method} & \textbf{Prec.} & \textbf{Rec.} & \textbf{F1} & \textbf{Acc.} \\
\midrule
\multirow{2}{*}{SPI}
  & Lu \textit{et al.}~\cite{lu2023} & ---   & ---   & ---   & 98.00 \\
  & FT-Pilot                         & 96.10 & 98.67 & 97.37 & 95.65 \\
\midrule
\multirow{2}{*}{I2C}
  & Lu \textit{et al.}~\cite{lu2023} & ---   & ---   & ---   & {\footnotesize 80.39--91.49} \\
  & FT-Pilot                         & 93.10 & 84.38 & 88.52 & 85.11 \\
\midrule
\multirow{2}{*}{XGE-MAC}
  & Lu \textit{et al.}~\cite{lu2023} & ---   & 98.83 & ---   & 96.67 \\
  & FT-Pilot                         & 95.77 & 99.27 & 97.49 & 96.73 \\
\midrule
\multirow{2}{*}{mriscv}
  & Lu \textit{et al.}~\cite{lu2023}$^{\dagger}$ & 97.74 & 97.74 & 97.74 & 98.67 \\
  & FT-Pilot                         & 98.53 & 100.00 & 99.26 & 98.84 \\
\bottomrule
\multicolumn{6}{@{}p{\linewidth}@{}}{\footnotesize $^{\dagger}$~\cite{lu2023} evaluates on RI5CY; we evaluate on mriscv. Both are RV32I cores.}
\end{tabular}
\end{table}

The experimental results in Table~\ref{tab:gnn-main} show that the proposed GraphSAGE-based predictor achieves strong classification performance across all four designs, and outperforms or remains comparable to the method of Lu et al.~\cite{lu2023} on most metrics. It is worth noting that our method performs vulnerability prediction directly at the pre-synthesis RTL level, whereas the method of Lu et al. operates at the netlist level; therefore, our approach avoids the additional synthesis-related processing flow. On mriscv, the model achieves an F1 score of 99.26\% and an accuracy of 98.84\%, indicating that the proposed method remains stable on a larger processor core. On XGE-MAC, our method is slightly better than that of Lu et al. on both comparable metrics, namely accuracy (96.73\% vs. 96.67\%) and recall (99.27\% vs. 98.83\%). On SPI, the proposed method also achieves high precision (96.10\%) and recall (98.67\%), with an F1 score of 97.37\%; although its accuracy is slightly lower than the 98.00\% reported by Lu et al., the overall performance remains at a high level. On I2C, the model attains an accuracy of 85.11\%. Since this design is relatively small, the prediction results are more easily affected by class distribution and sample size. Lu et al. also observed similar accuracy fluctuations on this circuit (80.39\%--91.49\%), and our result falls within the range reported in their study.

\begin{table}[htbp]
\centering
\caption{Time required for fault simulation versus GNN inference.}
\label{tab:gnn-time}
\small
\setlength{\tabcolsep}{8pt}
\begin{tabular}{ccc}
\toprule
\textbf{Circuit} & \textbf{\begin{tabular}[c]{@{}c@{}}Time required for\\ fault simulation\end{tabular}} & \textbf{\begin{tabular}[c]{@{}c@{}}Time required\\ for GNNs\end{tabular}} \\
\midrule
SPI     & 63.0\,s     & 10.52\,s \\
I2C     & 5410.4\,s   & 10.14\,s \\
XGE-MAC & 843.7\,s    & 24.06\,s \\
mriscv  & 10227.6\,s  & 19.24\,s \\
\bottomrule
\end{tabular}
\end{table}

Compared with the conventional fault-injection flow, the proposed GNN-based method offers a significant advantage in analysis efficiency. Table~\ref{tab:gnn-time} compares the time cost of fault-injection simulation with that of the proposed GNN-based approach for each of the four designs. Fault injection requires about 10{,}228 seconds (nearly 3 hours) on mriscv and about 5{,}410 seconds (roughly 1.5 hours) on I2C, whereas the GNN-based method completes within 25 seconds for all designs. Together with the classification results in Table~\ref{tab:gnn-main}, these results show that the proposed method can substantially reduce the time cost of vulnerability analysis while maintaining high accuracy, thereby supporting reliability evaluation at the early RTL design stage.
\subsection{LLM-Based RTL Hardening Results}

\begin{table*}[htbp]
\centering
\caption{End-to-end comparison on all 14 benchmark designs. Area in $\mu m^2$, Delay in ns, AO (area overhead) and Err in \%. ``---'' indicates that the data was not available for the corresponding design.}
\label{tab:main}
\small
\setlength{\tabcolsep}{3.5pt}
\begin{tabular}{l rrc rcrc rcrc rcrc}
\toprule
\multirow{2}{*}{\textbf{Design}} 
  & \multicolumn{3}{c}{\textbf{Unhardened}} 
  & \multicolumn{4}{c}{\textbf{Full TMR}} 
  & \multicolumn{4}{c}{\textbf{GNN+TMRG~\cite{TMRG}}} 
  & \multicolumn{4}{c}{\textbf{FT-Pilot}} \\
\cmidrule(lr){2-4} \cmidrule(lr){5-8} \cmidrule(lr){9-12} \cmidrule(lr){13-16}
 & Area & Delay & Err 
 & Area & AO & Delay & Err 
 & Area & AO & Delay & Err 
 & Area & AO & Delay & Err \\
\midrule
alu             & 173.4   & 0.506 & \textbf{12.28} & 321.3  & +85.29  & 0.506 & 0.00 & 343.1  & +97.87  & 0.506 & 2.90 & 315.5    & \textbf{+81.95}  & 0.506 & \textbf{2.90} \\
fifo            & 1226.0  & 0.399 & \textbf{46.10} & ---    & ---     & ---   & ---  & ---    & ---     & ---   & ---  & 2211.5   & \textbf{+80.38}  & 0.695 & \textbf{0.00} \\
fsm1            & 30.1    & 0.174 & \textbf{41.25} & 72.4   & +140.53 & 0.137 & 0.00 & 81.1   & +169.44 & 0.174 & 0.00 & 72.4     & \textbf{+140.53} & 0.111 & \textbf{0.00} \\
parallel2serial & 48.1    & 0.114 & \textbf{22.29} & 162.0  & +236.80 & 0.112 & 0.00 & 103.2  & +114.55 & 0.112 & 0.31 & 93.9     & \textbf{+95.22}  & 0.112 & \textbf{0.31} \\
serial2parallel & 155.1   & 0.208 & \textbf{29.26} & 497.2  & +220.57 & 0.208 & 0.00 & 311.0  & +100.52 & 0.208 & 3.83 & 297.7    & \textbf{+91.94}  & 0.483 & \textbf{3.83} \\
memory          & 33655.9 & 0.435 & \textbf{6.93}  & ---    & ---     & ---   & ---  & ---    & ---     & ---   & ---  & 54619.4  & \textbf{+62.29}  & 0.869 & \textbf{0.00} \\
\midrule
crc             & 364.2   & 0.200 & \textbf{31.04} & 1117.2 & +206.75 & 0.189 & 0.00 & 971.2  & +166.67 & 0.216 & 0.16 & 665.0    & \textbf{+82.59}  & 0.348 & \textbf{0.16} \\
gemm            & 600911.0& 1.747 & \textbf{4.90}  & ---    & ---     & ---   & ---  & ---    & ---     & ---   & ---  & 766571.0 & \textbf{+27.57}  & 1.931 & \textbf{0.18} \\
serial\_io      & 370.8   & 0.463 & \textbf{7.38}  & 984.7  & +165.56 & 0.556 & 0.00 & 519.8  & +40.18  & 0.504 & 1.91 & 504.1    & \textbf{+35.95}  & 0.627 & \textbf{1.91} \\
bus\_a          & 466.6   & 0.418 & \textbf{8.65}  & 1219.3 & +161.32 & 0.481 & 0.00 & 854.4  & +83.11  & 0.433 & 2.97 & 663.1    & \textbf{+42.11}  & 0.566 & \textbf{2.97} \\
uart            & 790.8   & 0.730 & \textbf{5.57}  & 2605.7 & +229.50 & 0.812 & 0.00 & 1595.2 & +101.72 & 0.781 & 0.18 & 1077.6   & \textbf{+36.27}  & 0.840 & \textbf{0.18} \\
\midrule
i2c\_master     & 1488.5  & 0.657 & \textbf{42.07} & 4134.4 & +177.76 & 0.739 & 0.00 & 3386.4 & +127.50 & 0.750 & 6.05 & 2751.0   & \textbf{+84.82}  & 0.722 & \textbf{6.05} \\
spi\_master     & 3350.8  & 1.064 & \textbf{10.25} & 8282.2 & +147.17 & 1.015 & 0.00 & 8032.4 & +139.72 & 1.021 & 2.53 & 5163.3   & \textbf{+54.09}  & 1.001 & \textbf{2.53} \\
mriscv          & 25604.1 & 1.236 & \textbf{6.20}  & 67332.0& +162.97 & 1.998 & 0.00 & 47815.6& +86.75  & 1.307 & 1.46 & 35788.2  & \textbf{+39.78}  & 1.694 & \textbf{1.46} \\
\bottomrule
\end{tabular}
\end{table*}
In this section, we compare the proposed framework against three baselines---Unhardened, Full TMR, and GNN+TMRG---on all 14 benchmark designs. The evaluation focuses on three aspects: hardware overhead, reliability under fault injection, and timing impact. All 14 designs pass the multi-level verification flow, including syntax checking, interface consistency checking, functional verification, and fault-injection-based reliability evaluation, confirming that the framework is applicable to RTL designs with varying scales and functional characteristics. Table~\ref{tab:main} reports the area, critical-path delay, and error rate under SEU fault injection for the four schemes. Among the 14 designs, Full TMR and GNN+TMRG baselines are available for 11 designs; the remaining three (fifo, memory, and gemm) are excluded from these two baselines. For storage-dominated designs such as fifo and memory, applying TMR to all registers is of limited practical value, since nearly all internal storage elements would be classified as registers and triplicated, resulting in roughly 200\% overhead without meaningful selectivity. Similarly, gemm contains large matrix storage arrays that dominate the design area, making uniform TMR impractical, the proposed framework instead applies an ABFT strategy to protect the computation results at a significantly lower cost. These three designs are included primarily to demonstrate the applicability of the proposed framework to diverse circuit types, where our method can select more efficient strategies such as ECC and ABFT in place of TMR.

Taking the original Unhardened design as the area baseline, Full TMR incurs an average area overhead of +176\% across the 11 designs where it is evaluated, because triple modular redundancy is uniformly applied to all registers; on parallel2serial and uart, the overhead exceeds +220\%, indicating the high cost of global hardening. By using the GNN predictor to prioritize critical registers for protection, GNN+TMRG reduces the average overhead to +112\%; however, the cost remains relatively high, mainly because this method can only apply TMR uniformly to all critical registers. In contrast, the proposed method achieves an average area overhead of only +71\% on the same 11-design subset, which is substantially lower than both Full TMR and GNN+TMRG. This advantage is particularly evident on datapath-dominated designs such as uart (+36\%), serial\_io (+36\%), and mriscv (+40\%). The main reason is that our framework can assign lower-cost correction strategies, such as Hamming or SECDED codes, to wide data registers, whereas GNN+TMRG is constrained by tool capability and can only employ a single TMR-based strategy. Across all 14 designs, the area overhead of the proposed method ranges from +28\% (gemm) to +141\% (fsm1). The highest overhead appears in fsm1, mainly because this design is extremely small and all of its registers are identified as critical, making the result close to Full TMR. In contrast, for medium- and large-scale designs, selective hardening demonstrates a more pronounced area advantage.

In terms of reliability, GNN+TMRG and the proposed method use the same GNN predictor and therefore protect exactly the same set of registers; the only difference lies in the subsequent strategy selection. However, the fault-tolerance strategies selected by our framework, together with the TMR strategy applied by GNN+TMRG, can all provide single-error correction under the SEU single-bit fault model. As a result, the two methods achieve identical error rates in the fault-injection experiments. As the upper bound of reliability, Full TMR reduces the error rate to 0\%. Meanwhile, both the proposed method and GNN+TMRG substantially reduce the error rate on all designs. Specifically, the error rates of fifo, fsm1, and memory are reduced to 0\%, while those of crc, uart, and gemm are reduced to below 0.2\%. Even for i2c\_master, which has the highest original error rate, the error rate decreases from 42.07\% to 6.05\% (an 85.6\% reduction). These residual errors mainly come from registers that are predicted by the GNN as non-critical and therefore remain unhardened.

As for timing overhead, the impact varies across design scales and hardening strategies. For small designs, the hardening logic generally has minimal impact on timing. On alu, the critical-path delay is identical across all four configurations, indicating that the added protection logic does not alter the critical path. On parallel2serial, the variation among configurations is within 0.002\,ns, which is attributable to synthesis noise rather than a meaningful timing penalty. A notable case is fsm1, where the proposed method achieves a shorter critical path (0.111\,ns) than the original (0.174\,ns); this can be attributed to improved code quality introduced during rewriting, such as adding explicit \texttt{default} branches in \texttt{case} statements, which enables the synthesizer to produce more efficient logic. For storage-dominated designs such as fifo and memory, the proposed framework selects ECC-based strategies, which inherently introduce encoding and decoding logic on the data path, resulting in noticeable delay increases; this is an intrinsic cost of these correction-capable strategies rather than a limitation of the framework itself. For medium- and large-scale designs, both the proposed method and GNN+TMRG introduce moderate timing overhead, and their overall delay impact remains at a comparable level.

Overall, while maintaining the same reliability as GNN+TMRG, the proposed method further reduces the average area overhead from +112\% to +71\% on the 11-design subset with complete baseline data. Compared with Full TMR, it achieves a more favorable cost--reliability tradeoff while sacrificing only a small amount of reliability.

Beyond hardware overhead and reliability, the proposed method also offers significant advantages in terms of workflow automation. As summarized in Table~\ref{tab:auto}, the conventional manual hardening flow requires designer involvement in all four stages. TMRG partially automates the rewriting stage but still requires manually prepared protection lists and constraint files, and its strategy is limited to TMR. GNN+TMRG further automates vulnerability identification by replacing manual analysis with a GNN predictor; however, strategy selection remains restricted to TMR, and the constraint files must still be manually derived from the GNN outputs.
\begin{table}[htbp]
\centering
\caption{Automation level comparison of hardening flows.}
\label{tab:auto}
\small
\setlength{\tabcolsep}{4pt}
\begin{tabular}{lcccc}
\toprule
\textbf{Flow} & \textbf{Vuln. Id.} & \textbf{Strategy} & \textbf{Rewrite} & \textbf{Verification} \\
\midrule
Manual        & Manual & Manual         & Manual    & Manual \\
TMRG          & Manual & TMR only       & Semi-Auto & Manual \\
GNN+TMRG      & GNN    & TMR only       & Semi-Auto & Manual \\
\textbf{FT-Pilot} & GNN   & Multi-strategy & LLM      & Auto \\
\bottomrule
\end{tabular}
\end{table}

 In contrast, the proposed method automates all of these key stages: the GNN automatically identifies vulnerable registers, the RTL Analyzer generates multi-strategy hardening schemes, and the RTL Rewriter completes RTL rewriting automatically with LLM assistance. Furthermore, the multi-level verification pipeline---including syntax checking, interface consistency checking, functional simulation, and fault injection---is also fully automated, enabling a closed-loop flow from vulnerability identification to verified hardened RTL without manual intervention.
\subsection{Ablation Study}
\begin{figure}[htbp]
\centering
\includegraphics[width=\linewidth]{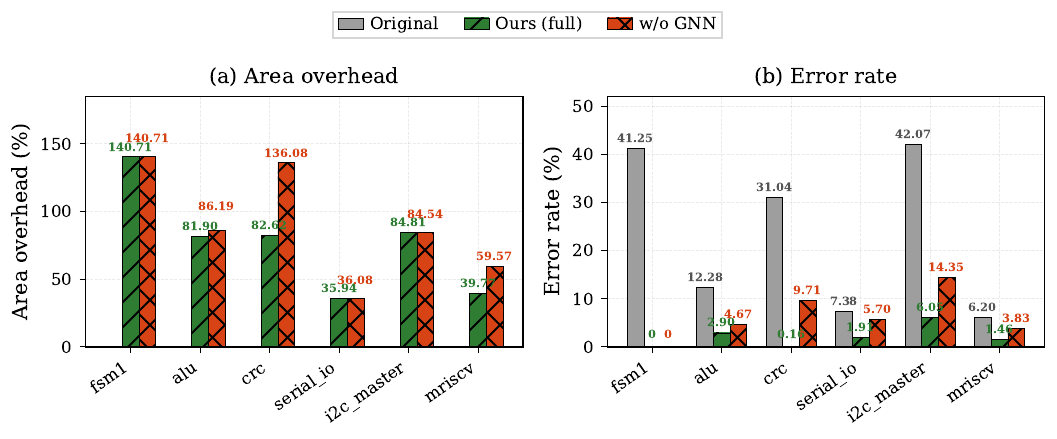}
\caption{Ablation study on GNN-based vulnerability prediction: area overhead and error rate comparison across six representative designs.}
\label{fig:w/o_gnn}
\end{figure}

To validate the contribution of the two core components in the proposed framework, we conduct ablation studies on two dimensions. For the GNN component, we evaluate six representative designs and compare the full method with a w/o GNN variant, in which the GNN-based vulnerability predictor is removed and the LLM autonomously selects which registers to harden based on its own code analysis, while the rewriting engine remains unchanged. For the RAG component, we evaluate all 14 designs and compare 
the full pipeline with a w/o RAG variant, in which the dual-knowledge-base retrieval module is removed so that no domain knowledge or code templates are injected during rewriting, while the Auto-Repair mechanism is still retained. The GNN ablation results are shown in 
Fig.~\ref{fig:w/o_gnn}, and the RAG ablation results are reported in Table~\ref{tab:ablation-rag}.

For the w/o GNN variant, since the LLM rewriting engine itself remains unchanged, the functional verification pass rate is unaffected. 
However, without GNN-guided vulnerability reports, the LLM tends to misallocate hardening resources to non-critical registers while leaving genuinely vulnerable ones unprotected. As shown in Fig.~\ref{fig:w/o_gnn}, 
the error rates of all six designs increase to varying degrees. Among them, crc exhibits the largest degradation, with its error rate rising from 0.16\% to 9.71\%, while i2c\_master increases from 6.05\% 
to 14.35\%. Meanwhile, the area overhead of the w/o GNN variant does not decrease---in fact, for designs such as crc (+136\% vs.\ +83\%) and mriscv (+60\% vs.\ +40\%), the overhead is notably higher, because the LLM without vulnerability guidance tends to 
over-protect low-risk registers. These results demonstrate that accurate vulnerability prediction is essential for achieving cost-effective selective hardening.

\begin{figure}[htbp]
\centering
\includegraphics[width=\linewidth]{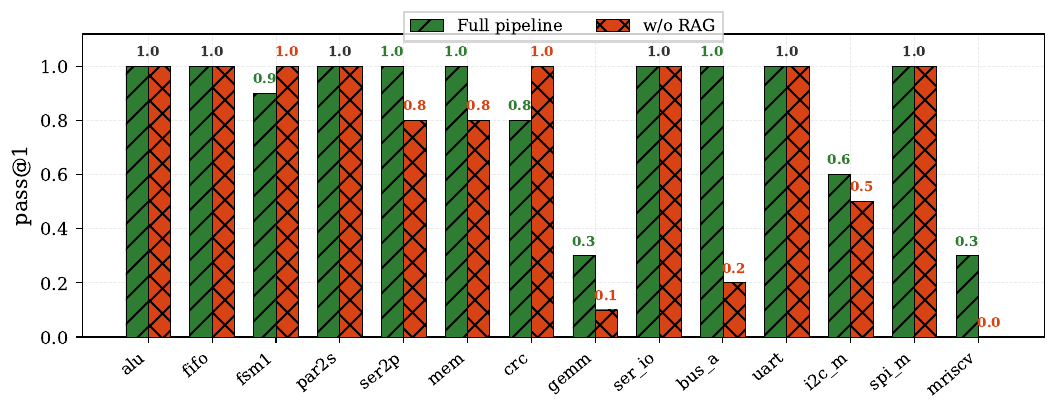}
\caption{Ablation study on RAG: per-design pass@1 for the full pipeline versus the w/o RAG variant.}
\label{fig:rag-ablation}
\end{figure}

\begin{table}[htbp]
\centering
\caption{Ablation study on RAG component, pass@k in \% (n=10, T=0.8).}
\label{tab:ablation-rag}
\small
\begin{tabular}{lcc}
\toprule
\textbf{Configuration} & \textbf{pass@1} & \textbf{pass@3} \\
\midrule
w/o RAG              & 74.29 & 83.93 \\
\textbf{Full pipeline} & \textbf{85.00} & \textbf{95.60} \\
\bottomrule
\end{tabular}
\end{table}
For the w/o RAG variant, the LLM loses access to domain knowledge and code templates during rewriting. As shown in Table~\ref{tab:ablation-rag} and Fig.~\ref{fig:rag-ablation}, this leads to a decrease in the average pass@1 from 85.00\% to 74.29\% and pass@3 from 95.60\% to 83.93\%. The impact is most pronounced on designs that require specialized hardening patterns: bus\_a drops from 100\% to 20\% in pass@1, and mriscv drops from 30\% to 0\%, indicating a complete failure to produce functionally correct hardened code without RAG support. gemm also decreases from 30\% to 10\%. Analysis of the failed samples reveals that, without RAG, the LLM is more prone to generating semantically defective hardening logic, such as incomplete TMR voter implementations, voting on stale register values, and check-bit misalignment in Hamming encoders---errors that correspond precisely to the failure modes addressed by the Semantic KB and Examples KB. For designs that pass verification under both configurations, the resulting area overhead and error rate remain comparable, confirming that RAG primarily improves the correctness and robustness of the rewriting process rather than the quality of already-correct hardening logic.
\subsection{Impact of LLM Selection}
\begin{figure}[htbp]
\centering
\includegraphics[width=\linewidth]{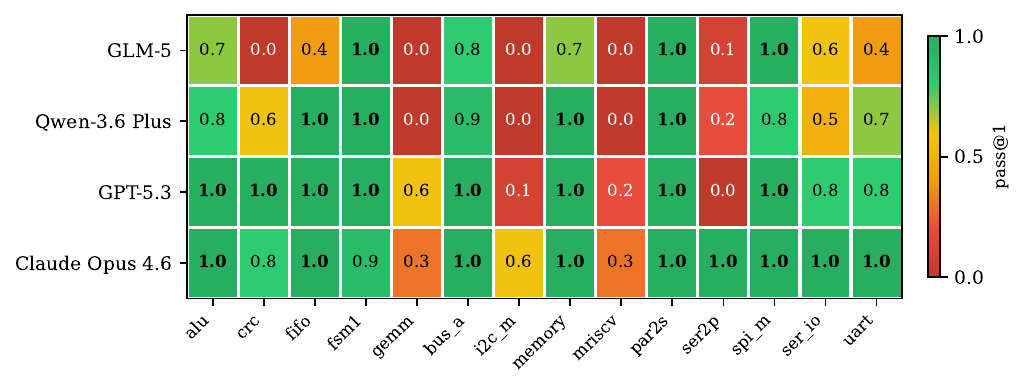}
\caption{Per-design pass@1 for four LLM backbones across 14 benchmark designs. Darker green indicates higher pass rates.}
\label{fig:llm-sensitivity}
\end{figure}

The rewriting capability of the proposed framework is fundamentally governed by the code comprehension and generation ability of the underlying LLM. To assess this sensitivity, we fix all other pipeline components and substitute four different LLMs—Claude Opus 4.6, GPT-5.3, GLM-5, and Qwen-3.6 Plus—as the backbone, running each of the 14 benchmark designs 10 times with a sampling temperature of 0.8. We adopt the pass@k metric to measure functional verification success rates. The quantitative results are summarized in Table~\ref{tab:llm-sens}, and a per-design breakdown is visualized as a heatmap in Fig.~\ref{fig:llm-sensitivity}.

The results confirm that LLM capability has a pronounced effect on rewriting success. Claude Opus 4.6 achieves the highest pass@1 of 85.00\% and pass@3 of 95.60\%, followed by GPT-5.3 at 75.00\% pass@1, while GLM-5 reaches only 47.86\%. The per-design heatmap in Fig.~\ref{fig:llm-sensitivity} further reveals that all four models reliably handle small, single-module designs such as fsm1 and parallel2serial, yet the gap widens sharply on multi-module designs (i2c\_master, mriscv) and computationally complex ones (gemm). Notably, each model exhibits distinct failure patterns---GPT-5.3 scores 0.0 on serial2parallel despite 1.0 on crc, whereas Claude Opus 4.6 is the only model to reach 1.0 on serial2parallel, indicating stronger cross-design generalization. Weaker models tend to introduce port mismatches, omit control conditions, or misplace hardening logic when confronted with multi-module, long-context rewriting tasks. These findings suggest that pairing the framework with a more capable LLM yields higher rewriting reliability, and that continued advances in LLM code generation are likely to translate directly into improved end-to-end automation.

\begin{table}[htbp]
\centering
\caption{LLM sensitivity study, pass@k in \% (14 designs, n=10, T=0.8).}
\label{tab:llm-sens}
\small
\begin{tabular}{lcc}
\toprule
\textbf{LLM Backbone} & \textbf{pass@1} & \textbf{pass@3} \\
\midrule
GLM-5                        & 47.86 & 63.69 \\
Qwen-3.6 Plus                & 60.71 & 74.35 \\
GPT-5.3                      & 75.00 & 84.29 \\
\textbf{Claude Opus 4.6}     & \textbf{85.00} & \textbf{95.60} \\
\bottomrule
\end{tabular}
\end{table}

\section{Discussion}
The experiments on 14 benchmark circuits demonstrate that FT-Pilot can automate the entire flow from vulnerability prediction to fault-tolerant code generation. Nevertheless, several limitations merit further discussion.
\subsection{Dependence of the framework on LLM capability}
As demonstrated in Section IV-E, the rewriting success rate depends substantially on the code comprehension and generation ability of the underlying LLM. Among the four models evaluated, Claude Opus 4.6 achieves the highest average pass@1 of 85.00\%, whereas GLM-5 reaches only 47.86\%, with the performance gap becoming particularly pronounced on multi-module and structurally complex designs. This indicates that the overall effectiveness of the proposed framework is, to a considerable extent, bounded by the current capability of large language models. In addition to cross-model variation, the inherent stochasticity of LLM generation introduces within-model variability: since sampling is performed with a temperature of T = 0.8, multiple independent runs on the same design may yield hardened code of differing quality. The gap between pass@1 and pass@3 (85.00\% vs. 95.60\% for Claude Opus 4.6) confirms that, for certain complex designs, a single sample may not succeed, yet drawing a small number of additional candidates substantially raises the probability of obtaining a correct result. For industrial deployment, reducing this variability and improving single-shot reliability remain important challenges. Nevertheless, as LLMs continue to advance in code generation, such progress is expected to directly translate into higher rewriting quality within the framework without requiring architectural modifications.
\subsection{Generalizability of vulnerability prediction and scope of fault models}
The current GNN predictor is trained on a per-design basis, meaning that for each new, previously unseen design, fault-injection simulation is still required to generate training labels, cross-design generalization has not yet been achieved. Since this is fundamentally a model generalization problem, we leave its in-depth investigation to future work. In addition, the experiments in this paper consider only the SEU fault model. As technology nodes continue to shrink, the probability of multi-bit upsets (MBUs) increases, and existing fault-tolerance strategies such as Hamming(7,4) encoding and standard TMR may no longer provide adequate protection. Extending the GNN predictor to generalize across unseen designs, and broadening the framework to accommodate more complex fault scenarios, represent important directions for future research.
\subsection{Framework scalability and broader applicability}
The largest benchmark circuit evaluated in this work, mriscv, comprises approximately 2,665 lines of code. Designs with significantly larger register counts, such as XGE-MAC (used in our GNN evaluation but excluded from LLM rewriting due to its scale), already exceed the practical context limits of current LLMs. Industrial designs are considerably larger, and both the limited context window of LLMs and the retrieval precision of RAG may become bottlenecks when the framework is applied to such designs. The ablation study further highlights the critical role of the knowledge base: removing RAG causes the average pass@1 to drop from 85.00\% to 74.29\% and the average pass@3 from 95.60\% to 83.93\%. However, the strategy entries in the current knowledge base are manually curated, and incorporating new fault-tolerance techniques requires ongoing human effort. These scalability challenges may be addressed in future work by decomposing large designs along their module hierarchy and by automatically extracting strategy knowledge from published literature and design specifications. From a broader perspective, the experimental results of this work suggest that LLMs exhibit considerable potential for understanding the functional semantics of RTL code and performing targeted code transformations accordingly. In principle, this paradigm of automated, semantics-driven code transformation could extend beyond fault-tolerance hardening to other RTL transformation tasks that require design-level semantic knowledge, such as power optimization and timing closure, provided that rigorous multi-level verification is employed to ensure the correctness of the generated results.
\section{Conclusion}
This paper presents FT-Pilot, a GNN-guided LLM framework for automated RTL-level soft-error hardening. A GNN operating on AIG-based graphs predicts register-level vulnerability directly at the RTL level, while an LLM-driven rewriting engine, supported by dual-knowledge-base RAG and an automatic repair mechanism, performs strategy selection and fault-tolerant code transformation with multi-level verification. Experiments on 14 benchmark circuits demonstrate that the GNN predictor achieves 85\%--99\% accuracy with inference times orders of magnitude faster than fault injection, and the rewriting engine produces verified hardened code on all designs while reducing average area overhead from +112\% to +71\% compared with GNN+TMRG at the same reliability level.

\section*{Acknowledgment}
Claude (Anthropic) was employed exclusively for syntactic and grammatical polishing to ensure the manuscript meets academic standards; the AI was not involved in data analysis or the development of the research findings.

\bibliographystyle{IEEEtran}   

\bibliography{references}

@ARTICLE{lu2023,
  author={Lu, Li and Chen, Junchao and Ulbricht, Markus and Krstic, Milos},
  journal={IEEE Transactions on Computer-Aided Design of Integrated Circuits and Systems}, 
  title={Toward Critical Flip-Flop Identification for Soft-Error Tolerance With Graph Neural Networks}, 
  year={2024},
  volume={43},
  number={4},
  pages={1135-1148},
  doi={10.1109/TCAD.2023.3331968}}

@article{lyons1962tmr,
  author  = {R. E. Lyons and W. Vanderkulk},
  title   = {The use of triple-modular redundancy to improve computer reliability},
  journal = {IBM Journal of Research and Development},
  volume  = {6},
  number  = {2},
  pages   = {200--209},
  year    = {1962}
}

@ARTICLE{abft,
  author={Kuang-Hua Huang and Abraham, Jacob A.},
  journal={IEEE Transactions on Computers}, 
  title={Algorithm-Based Fault Tolerance for Matrix Operations}, 
  year={1984},
  volume={C-33},
  number={6},
  pages={518-528},
  doi={10.1109/TC.1984.1676475}}

@ARTICLE{bubblerazor,
  author={Fojtik, Matthew and Fick, David and Kim, Yejoong and Pinckney, Nathaniel and Harris, David Money and Blaauw, David and Sylvester, Dennis},
  journal={IEEE Journal of Solid-State Circuits}, 
  title={Bubble Razor: Eliminating Timing Margins in an ARM Cortex-M3 Processor in 45 nm CMOS Using Architecturally Independent Error Detection and Correction}, 
  year={2013},
  volume={48},
  number={1},
  pages={66-81},
  doi={10.1109/JSSC.2012.2220912}}

@INPROCEEDINGS{razor,
  author={Ernst, D. and Nam Sung Kim and Das, S. and Pant, S. and Rao, R. and Toan Pham and Ziesler, C. and Blaauw, D. and Austin, T. and Flautner, K. and Mudge, T.},
  booktitle={Proceedings. 36th Annual IEEE/ACM International Symposium on Microarchitecture, 2003. MICRO-36.}, 
  title={Razor: a low-power pipeline based on circuit-level timing speculation}, 
  year={2003},
  volume={},
  number={},
  pages={7-18},
  doi={10.1109/MICRO.2003.1253179}}

@inproceedings{tang2025eraser,
  author    = {Tang, Jiaping and Mu, Jianan and Liu, Silin and
               Liu, Zizhen and Gu, Feng and Zhang, Xinyu and
               Wang, Leyan and Liang, Shenwen and Ye, Jing and
               Li, Huawei and Li, Xiaowei},
  title     = {{ERASER}: Efficient {RTL} Fault Simulation Framework
               with Trimmed Execution Redundancy},
  booktitle = {Proc. Design, Automation \& Test in Europe (DATE)},
  pages     = {1--7},
  year      = {2025},
  doi       = {10.23919/DATE64628.2025.10993024}
}

@inproceedings{li2020exploring,
  title={Exploring a bayesian optimization framework compatible with digital standard flow for soft-error-tolerant circuit},
  author={Li, Yan and Zeng, Xiaoyoung and Gao, Zhengqi and Lin, Liyu and Tao, Jun and Han, Jun and Cheng, Xu and Tahoori, Mehdi and Zeng, Xiaoyang},
  booktitle={2020 57th ACM/IEEE Design Automation Conference (DAC)},
  pages={1--6},
  year={2020},
  organization={IEEE}
}

@INPROCEEDINGS{lange2019ml,
  author={Lange, Thomas and Balakrishnan, Aneesh and Glorieux, Maximilien and Alexandrescu, Dan and Sterpone, Luca},
  booktitle={2019 IEEE 25th International Symposium on On-Line Testing and Robust System Design (IOLTS)}, 
  title={Machine Learning to Tackle the Challenges of Transient and Soft Errors in Complex Circuits}, 
  year={2019},
  volume={},
  number={},
  pages={7-14},
  doi={10.1109/IOLTS.2019.8854423}}

@INPROCEEDINGS{das2024gcn,
  author={Das, Sanjay and Kundu, Shamik and Madhusoodhanan, Pooja and Pillai, Prasanth Viswanathan and Parekhji, Rubin and Raha, Arnab and Banerjee, Suvadeep and Natarajan, Suriyaprakash and Basu, Kanad},
  booktitle={2024 61st ACM/IEEE Design Automation Conference (DAC)}, 
  title={Graph Learning-based Fault Criticality Analysis for Enhancing Functional Safety of E/E Systems}, 
  year={2024},
  volume={},
  number={},
  pages={1-6},
  doi={}}

@INPROCEEDINGS{autovcoder,
  author={Gao, Mingzhe and Zhao, Jieru and Lin, Zhe and Ding, Wenchao and Hou, Xiaofeng and Feng, Yu and Li, Chao and Guo, Minyi},
  booktitle={2024 IEEE 42nd International Conference on Computer Design (ICCD)}, 
  title={AutoVCoder: A Systematic Framework for Automated Verilog Code Generation using LLMs}, 
  year={2024},
  volume={},
  number={},
  pages={162-169},
  doi={10.1109/ICCD63220.2024.00033}}

@INPROCEEDINGS{mage,
  author={Zhao, Yujie and Zhang, Hejia and Huang, Hanxian and Yu, Zhongming and Zhao, Jishen},
  booktitle={2025 62nd ACM/IEEE Design Automation Conference (DAC)}, 
  title={MAGE: A Multi-Agent Engine for Automated RTL Code Generation}, 
  year={2025},
  volume={},
  number={},
  pages={1-7},
  doi={10.1109/DAC63849.2025.11133191}}

@INPROCEEDINGS{rtlcoder,
  author={Liu, Shang and Fang, Wenji and Lu, Yao and Zhang, Qijun and Zhang, Hongce and Xie, Zhiyao},
  booktitle={2024 IEEE LLM Aided Design Workshop (LAD)}, 
  title={RTLCoder: Outperforming GPT-3.5 in Design RTL Generation with Our Open-Source Dataset and Lightweight Solution}, 
  year={2024},
  volume={},
  number={},
  pages={1-5},
  doi={10.1109/LAD62341.2024.10691788}}

@INPROCEEDINGS{hlspilot,
  author={Xiong, Chenwei and Liu, Cheng and Li, Huawei and Li, Xiaowei},
  booktitle={2024 ACM/IEEE International Conference On Computer Aided Design (ICCAD)}, 
  title={HLSPilot: LLM-Based High-Level Synthesis}, 
  year={2024},
  volume={},
  number={},
  pages={1-9},
  doi={}}

@INPROCEEDINGS{llm4gv,
  author={Zou, Dingyang and Zhang, Gaoche and Sun, Kairui and Wen, Zhe and Wang, Meiqi and Wang, Zhongfeng},
  booktitle={2025 Design, Automation \& Test in Europe Conference (DATE)}, 
  title={LLM4GV: An LLM-Based Flexible Performance-Aware Framework for GEMM Verilog Generation}, 
  year={2025},
  volume={},
  number={},
  pages={1-2},
  doi={10.23919/DATE64628.2025.10992751}}

@INPROCEEDINGS{rtlrewriter,
  author={Yao, Xufeng and Wang, Yiwen and Li, Xing and Lian, Yingzhao and Chen, Ran and Chen, Lei and Yuan, Mingxuan and Xu, Hong and Yu, Bei},
  booktitle={2024 ACM/IEEE International Conference On Computer Aided Design (ICCAD)}, 
  title={RTLRewriter: Methodologies for Large Models Aided RTL Code Optimization}, 
  year={2024},
  volume={},
  number={},
  pages={1-7},
  doi={10.1145/3676536.3676775}}

@INPROCEEDINGS{selfhwdebug,
  author={Akyash, Mohammad and Kamali, Hadi Mardani},
  booktitle={2024 IEEE Computer Society Annual Symposium on VLSI (ISVLSI)}, 
  title={Self-HWDebug: Automation of LLM Self-Instructing for Hardware Security Verification}, 
  year={2024},
  volume={},
  number={},
  pages={391-396},
  doi={10.1109/ISVLSI61997.2024.00077}}

@INPROCEEDINGS{assertion,
  author={Pulavarthi, Vaishnavi and Nandal, Deeksha and Dan, Soham and Pal, Debjit},
  booktitle={2025 Design, Automation \& Test in Europe Conference (DATE)}, 
  title={Are LLMs Ready for Practical Adoption for Assertion Generation?}, 
  year={2025},
  volume={},
  number={},
  pages={1-7},
  doi={10.23919/DATE64628.2025.10992817}}

@ARTICLE{chateda,
  author={Wu, Haoyuan and He, Zhuolun and Zhang, Xinyun and Yao, Xufeng and Zheng, Su and Zheng, Haisheng and Yu, Bei},
  journal={IEEE Transactions on Computer-Aided Design of Integrated Circuits and Systems}, 
  title={ChatEDA: A Large Language Model Powered Autonomous Agent for EDA}, 
  year={2024},
  volume={43},
  number={10},
  pages={3184-3197},
  doi={10.1109/TCAD.2024.3383347}}

@ARTICLE{softerror,
  author={Baumann, R.C.},
  journal={IEEE Transactions on Device and Materials Reliability}, 
  title={Radiation-induced soft errors in advanced semiconductor technologies}, 
  year={2005},
  volume={5},
  number={3},
  pages={305-316},
  doi={10.1109/TDMR.2005.853449}}

@standard{iso26262,
  title        = {{ISO} 26262: Road Vehicles---Functional Safety},
  organization = {International Organization for Standardization},
  year         = {2018}
}

@ARTICLE{selectivehardening,
  author={Polian, Ilia and Hayes, John P.},
  journal={IEEE Design \& Test of Computers}, 
  title={Selective Hardening: Toward Cost-Effective Error Tolerance}, 
  year={2011},
  volume={28},
  number={3},
  pages={54-63},
  doi={10.1109/MDT.2010.120}}

@INPROCEEDINGS{Balakrishnan,
  author={Balakrishnan, Aneesh and Lange, Thomas and Glorieux, Maximilien and Alexandrescu, Dan and Jenihhin, Maksim},
  booktitle={2019 NASA/ESA Conference on Adaptive Hardware and Systems (AHS)}, 
  title={Modeling Gate-Level Abstraction Hierarchy Using Graph Convolutional Neural Networks to Predict Functional De-Rating Factors}, 
  year={2019},
  volume={},
  number={},
  pages={72-78},
  doi={10.1109/AHS.2019.00007}}

@article{ziade2004survey,
  title={A survey on fault injection techniques},
  author={Ziade, Haissam and Ayoubi, Rafic A and Velazco, Raoul and others},
  journal={Int. Arab J. Inf. Technol.},
  volume={1},
  number={2},
  pages={171--186},
  year={2004}
}

@article{TMRG,
  author={S. Kulis},
  title={Single Event Effects mitigation with TMRG tool},
  journal={Journal of Instrumentation},
  volume={12},
  number={01},
  pages={C01082},
  url={http://stacks.iop.org/1748-0221/12/i=01/a=C01082},
  year={2017}
}

@article{zhang2006soft,
  title={Soft-error-rate-analysis (SERA) methodology},
  author={Zhang, Ming and Shanbhag, Naresh R},
  journal={IEEE Transactions on Computer-Aided Design of Integrated Circuits and Systems},
  volume={25},
  number={10},
  pages={2140--2155},
  year={2006},
  publisher={IEEE}
}

@article{hamming1950error,
  title={Error detecting and error correcting codes},
  author={Hamming, Richard W},
  journal={The Bell system technical journal},
  volume={29},
  number={2},
  pages={147--160},
  year={1950},
  publisher={Nokia Bell Labs}
}

@ARTICLE{ml,
  author={{L. Lu, J. Chen, M. Ulbricht, and M. Krstic}},
  journal={IEEE Transactions on Circuits and Systems I: Regular Papers}, 
  title={Machine Learning Methodologies to Predict the Results of Simulation-Based Fault Injection}, 
  year={2024},
  volume={71},
  number={5},
  pages={1978-1991},
  doi={10.1109/TCSI.2024.3349928}}

@inproceedings{wolf2013yosys,
  title={Yosys-a free verilog synthesis suite},
  author={Wolf, Clifford and Glaser, Johann and Kepler, Johannes},
  booktitle={Proceedings of the 21st Austrian Workshop on Microelectronics (Austrochip)},
  volume={97},
  pages={1--6},
  year={2013}
}

@inproceedings{mukherjee2005soft,
  title={The soft error problem: An architectural perspective},
  author={Mukherjee, Shubhendu S and Emer, Joel and Reinhardt, Steven K},
  booktitle={11th International Symposium on High-Performance Computer Architecture},
  pages={243--247},
  year={2005},
  organization={IEEE}
}

@INPROCEEDINGS{fsm,
  author={Cassel, M. and Lima, F.},
  booktitle={12th IEEE International On-Line Testing Symposium (IOLTS'06)}, 
  title={Evaluating one-hot encoding finite state machines for SEU reliability in SRAM-based FPGAs}, 
  year={2006},
  volume={},
  number={},
  pages={6 pp.-},
  doi={10.1109/IOLTS.2006.32}}

@article{graphsage,
  title={Inductive representation learning on large graphs},
  author={Hamilton, Will and Ying, Zhitao and Leskovec, Jure},
  journal={Advances in neural information processing systems},
  volume={30},
  year={2017}
}

@article{rag,
  title={Retrieval-augmented generation for knowledge-intensive nlp tasks},
  author={Lewis, Patrick and Perez, Ethan and Piktus, Aleksandra and Petroni, Fabio and Karpukhin, Vladimir and Goyal, Naman and K{\"u}ttler, Heinrich and Lewis, Mike and Yih, Wen-tau and Rockt{\"a}schel, Tim and others},
  journal={Advances in neural information processing systems},
  volume={33},
  pages={9459--9474},
  year={2020}
}

@misc{synopsys_z01x,
  author       = {{Synopsys Inc.}},
  title        = {{VC Z01X} Functional Safety Simulator},
  howpublished = {\url{https://www.synopsys.com/verification/simulation/vc-z01x.html}},
  year         = {2023}
}

@article{cot,
  title={Chain-of-thought prompting elicits reasoning in large language models},
  author={Wei, Jason and Wang, Xuezhi and Schuurmans, Dale and Bosma, Maarten and Xia, Fei and Chi, Ed and Le, Quoc V and Zhou, Denny and others},
  journal={Advances in neural information processing systems},
  volume={35},
  pages={24824--24837},
  year={2022}
}

@inproceedings{rtllm_benchmark,
  title={Rtllm: An open-source benchmark for design rtl generation with large language model},
  author={Lu, Yao and Liu, Shang and Zhang, Qijun and Xie, Zhiyao},
  booktitle={2024 29th Asia and South Pacific Design Automation Conference (ASP-DAC)},
  pages={722--727},
  year={2024},
  organization={IEEE}
}

@misc{opencores,
  author       = {{OpenCores}},
  title        = {Open{C}ores---Open Source {IP} Cores},
  howpublished = {\url{https://opencores.org}},
  year         = {2023}
}

@misc{mriscv,
  author       = {{OnChipUIS}},
  title        = {{mriscvcore}: A {RISC-V} Processor Core},
  howpublished = {\url{https://github.com/onchipuis/mriscvcore}},
  year         = {2023}
}

@inproceedings{pytorchgeometric,
  author    = {Fey, Matthias and Lenssen, Jan Eric},
  title     = {Fast Graph Representation Learning with {PyTorch Geometric}},
  booktitle = {Proc. ICLR Workshop on Representation Learning on
               Graphs and Manifolds},
  year      = {2019}
}

@article{passk,
  title={Evaluating large language models trained on code},
  author={Chen, Mark and Tworek, Jerry and Jun, Heewoo and Yuan, Qiming and Pinto, Henrique Ponde De Oliveira and Kaplan, Jared and Edwards, Harri and Burda, Yuri and Joseph, Nicholas and Brockman, Greg and others},
  journal={arXiv preprint arXiv:2107.03374},
  year={2021}
}

@misc{williams2002iverilog,
  author       = {Williams, Stephen},
  title        = {Icarus Verilog},
  howpublished = {\url{https://github.com/steveicarus/iverilog}},
  year         = {2002}
}

@inproceedings{stine2007nangate,
  author    = {Stine, James E. and others},
  title     = {{FreePDK}: An Open-Source Variation-Aware Design Kit},
  booktitle = {Proc. IEEE Int. Conf. Microelectronic Systems Education (MSE)},
  pages     = {173--174},
  year      = {2007},
  doi       = {10.1109/MSE.2007.44}
}

@INPROCEEDINGS{features,
  author={Lange, Thomas and Balakrishnan, Aneesh and Glorieux, Maximilien and Alexandrescu, Dan and Sterpone, Luca},
  booktitle={2019 49th Annual IEEE/IFIP International Conference on Dependable Systems and Networks – Supplemental Volume (DSN-S)}, 
  title={On the Estimation of Complex Circuits Functional Failure Rate by Machine Learning Techniques}, 
  year={2019},
  volume={},
  number={},
  pages={35-41},
  doi={10.1109/DSN-S.2019.00021}}

\end{document}